A Computational Model of Infant Learning and Reasoning with Probabilities

Thomas R. Shultz,
Department of Psychology and School of Computer Science, McGill University

Ardavan S. Nobandegani
Department of Psychology and Department of Electrical & Computer Engineering, McGill University

Correspondence concerning this article should be addressed to Thomas R. Shultz, Department of Psychology, McGill University, 2001 Av. McGill College, Montreal, QC, Canada H3A 1G1. This research was supported in part by a grant from the Natural Sciences and Engineering Research Council of Canada to TRS.
Contact: thomas.shultz@mcgill.ca

Abstract

Recent experiments reveal that 6- to 12-month-old infants can learn probabilities and reason with them. In this work, we present a novel computational system called Neural Probability Learner and Sampler (NPLS) that learns and reasons with probabilities, providing a computationally sufficient mechanism to explain infant probabilistic learning and inference. In 24 computer simulations, NPLS simulations show how probability distributions can emerge naturally from neural-network learning of event sequences, providing a novel explanation of infant probabilistic learning and reasoning. Three mathematical proofs show how and why NPLS simulates the infant results so accurately. The results are situated in relation to seven other active research lines. This work provides an effective way to integrate Bayesian and neural-network approaches to cognition.

Keywords: infants, probabilistic learning and inference, neural networks, sampling

A Computational Model of Infant Learning and Reasoning with Probabilities

**Literature Review**

Succeeding in an uncertain world requires probabilistic reasoning: the ability to compute and act on relevant probabilities. Classical developmental psychological research suggested that the ability to perform basic forms of probabilistic reasoning does not develop in humans until around seven years of age (Piaget & Inhelder, 1975). However, a series of recent experiments has established that preverbal human infants can compute and reason with probabilities (Denison, Reed, & Xu, 2013; Xu & Garcia, 2008), even using them to guide their intentional behavior (Denison & Xu, 2010, 2014). Among the interesting findings are that 8-month-old infants can effectively generalize between populations and samples (Xu & Garcia, 2008), this ability emerges between 4.5 and 6 months of age (Denison et al., 2013), and infants can appropriately act on their probabilistic knowledge, across a wide range of task parameterizations (Denison & Xu, 2010, 2014).

These empirical results are impressive and it has been unclear how such young, preverbal infants accomplish these probabilistic tasks. Here, we develop, test, and analyze a new neural-network system, called Neural Probability Learner and Sampler (NPLS), applying it to this series of infant results. Our results accurately simulate the infant findings and make several testable predictions. Deeper literature reviews accompany each of our simulation results sections. In the Discussion section, we situate our results among seven other active literatures on probabilistic and quantitative reasoning.

For the last 2.5 decades, there has been an active literature on the computational modeling of early psychological development with artificial neural networks, well-documented in both review articles (Elman, 2005; Mareschal, 2010; Munakata & McClelland, 2003; Shultz, 2012, 2013, 2017; Shultz & Sirois, 2008; Westermann, Sirois, Shultz, & Mareschal, 2006) and books (Elman et al., 1996; Mareschal et al., 2007; Shultz, 2003). This paper contributes to that research, with implications for further integration with Bayesian approaches.

## Methods

Learning in NPLS is based on a constructive neural-learning algorithm called Sibling-Descendant Cascade-Correlation (SDCC) that builds the interior layers of a neural network during learning (Baluja & Fahlman, 1994). Probability distributions are estimated by a network's output activations. Greater infant surprise at unexpected than expected outcomes is simulated as error in network predictions. Both network error and infant surprise are characterized by the discrepancy between what is expected and what actually occurs (Althaus, Gliozzi, Mayor, & Plunkett, 2020; Oakes, Madole, & Cohen, 1991; Shultz & Cohen, 2004). Several enhancements of SDCC are required to cope with learning and using probability distributions. Reasoning in NPLS is performed by a Markov-chain Monte Carlo sampling algorithm (MCMC) that simulates infant sampling from the learned probability distributions, thereby explaining infant search directions for preferred objects.

### Neural-network Learning

An artificial neural network includes a set of units (roughly representing neurons or groups of neurons) that can be variously active, where activity is an average firing rate over a time period, and connection weights (roughly representing synapses), whose initial values are small and random, but which change during learning in order to reduce network error.

SDCC networks are deterministic, discriminative, feed-forward, networks that learn from examples by reducing overall prediction error (Baluja & Fahlman, 1994). They learn to discriminate categories of the examples they are trained on. Activations are passed forward: from input units that describe examples to layers of hidden units that transform inputs into more abstract representations, and finally to output units describing a response to some particular input. Network output can be considered a prediction or expectation of what will happen at the output layer, while target output represents what actually happens. During learning (in what is called output phase), connection weights are altered so that overall network error is reduced:

$$E = \sum_o \sum_p (A_{op} - T_{op})^2 \qquad \text{(Equation 1)}$$

where $E$ is sum-of-squared error, $A$ is the actual output activation for unit $o$ and pattern $p$, and $T$ is the target output activation for this unit and pattern.

SDCC training starts with a two-layer network (i.e., only the input and the output layer), and then recruits hidden units one at a time, as needed, to solve the problem being learned. The SDCC algorithm constructs its own network topology, as opposed to being hand designed by a programmer. In what is called input phase, input weights to candidate hidden units are trained to increase the covariation of candidate hidden unit output activation with current network error. When these covariations stop increasing, the highest correlating unit is then installed either on the current highest layer of hidden units or on its own higher layer, depending on which yields the better absolute covariation with existing error. Because input weights to each recruited hidden unit are frozen as the unit is installed, the SDCC algorithm adjusts weights only one layer at a time, thus never having to (unrealistically) propagate error signals backwards through the network. The function to maximize in input phase is a covariance between candidate-hidden-unit activation and network error:

$$C = \frac{\sum_o |\sum_p (h_p - \langle h \rangle)(e_{op} - \langle e_o \rangle)|}{\sum_o \sum_p (e_{op} - \langle e_o \rangle)^2} \qquad \text{(Equation 2)}$$

where $C$ is the covariance, $h_p$ is the activation of the candidate hidden unit for pattern $p$, $\langle h \rangle$ is the mean activation of the candidate hidden unit for all patterns, $e_{op}$ is the residual error at output $o$ for pattern $p$, and $\langle e_o \rangle$ is the mean residual error at output $o$ for all the training patterns. SDCC networks used here utilize an asymmetric sigmoid activation function:

$$y_i = \frac{1}{1 + e^{-x_i}} \qquad \text{(Equation 3)}$$

where $y$ is the receiving unit $i$'s output, $x$ is the net input to unit $i$, and $e$ is the exponential function. Thus, output unit activation ranges from a floor of 0 to a ceiling of 1, just as probabilities do.

SDCC has previously been used to simulate many deterministic phenomena in learning and cognitive development (Shultz, 2003, 2012, 2017), notably including several simulations on infant habituation of attention (Shultz, 2010; Shultz & Bale, 2001, 2006; Shultz & Cohen, 2004).

As constructive algorithms, CC and SDCC automatically construct the network interior by recruiting as many hidden units as required to learn the training patterns, thus capturing both development (via unit recruitment) and learning (via weight adjustment). The main difference between CC and SDCC is in potential network topology. Every hidden unit in CC is installed on its own layer, while SDCC dynamically decides whether to install each recruited hidden unit on the current highest layer or on its own higher layer. Four of the eight candidate hidden units are destined to be potential siblings and the other four are destined to be potential descendants of the current highest layer of hidden units. During each recruitment phase, the input-side weights of the each of the eight candidates are adjusted in order to increase the covariance between the candidate's activation and overall network error. When increases in covariances stagnate, the candidate with the highest covariance with network error is installed into its designated layer in the network, while the other seven candidates are discarded.

Simulation coverage of psychological data by these constructive networks is typically better than that achieved by symbolic rule models or static neural networks (Shultz, 2003, 2012, 2017). Constructive networks are particularly adept at simulating developmental stages because they do as well as they can with their current computational power. They start small, and gradually increase their computational power only as needed. The neural plausibility of these constructive networks is considerable and is discussed elsewhere (Shultz, 2003, 2017).

However, like other feed-forward, deterministic, discriminative neural networks, SDCC had limitations that precluded its immediate application to modeling Bayesian-inspired experiments. First, because of its determinism, it remained unsatisfied with the high error of probabilistic outcomes, recruiting new and useless hidden units ad infinitum. Second, it could not probabilistically generate novel examples from the categories it had learned.

The first problem was solved by allowing SDCC to track its own progress, measured in terms of error reduction over learning cycles. It already had the capacity to monitor progress within both input and output phases, using threshold and patience parameters. In output phases, SDCC adjusts connection weights to reduce error. When error reduction stagnates, it switches to input phase to recruit a new hidden unit, adjusting weights entering candidate units to increase the correlation between their activations and network error. In each of these phases, stagnation is detected when there is no progress greater than the threshold parameter for the number of training epochs specified by the patience parameter. This scheme was extended by adding an encompassing loop that uses its own threshold and patience parameters to monitor progress over learning cycles, where each learning cycle is defined as an input phase and the next output phase (Shultz & Doty, 2014). This learning cessation mechanism allows SDCC to stop learning when it no longer improves in reducing error over learning cycles. Learning threshold and patience parameters control how much learning is done. There is evidence that even infants give up on learning when they cease to make progress (Gerken, Balcomb, & Minton, 2011).

With learning cessation, SDCC can learn any (unnormalized) multivariate probability distribution from examples that specify whether or not an output occurs in the presence of a particular input (Kharratzadeh & Shultz, 2016). This turns out to be an efficient way to learn and represent probability distributions because the output activations match ground-truth probabilities at that point. In contrast, most Bayesian-inspired psychological research does not explicitly specify how probability distributions are learned in a mechanistic, brain-like way. Overcoming the second limitation, probabilistically generating examples, requires the use of an MCMC sampling algorithm, as described in the next subsection.

For the learning simulations presented here, networks are trained on multiples of two samples illustrating the frequency ratio of interest, e.g., 4:1 versus 1:4. Each network starts with an input unit, an output unit, and no hidden units. Network inputs of 1 or 2 arbitrarily represent the identities of two categories of training stimuli, here often representing a container holding objects of two particular colors. The occurrence of an event, e.g., appearance of an object of a particular color is represented by a target output of 1, while its absence is represented by a target output of 0. A pattern of 5 objects thus requires 5 input-output training pairs. When a particular color is in a 4:1 majority, 4 pairs would have an output of 1 and 1 pair would have an output of 0 (see Table 1). When a color is in a 1:4 minority, 1 pair has an output of 1 and 4 others an output of 0. The ordering of these input-output pairs does not matter, because SDCC training is done in batch mode, wherein weights are adjusted after one presentation of all the training patterns.

Each example pair is thus deterministic, with the aggregate pattern across pairs appearing to be probabilistic. For a given input, sometimes the output event occurs, and other times it does not. Test samples are created in an analogous fashion, with output values that would be either expected or surprising.

This coding of target outputs directly reflects the visual information given to infants in each experiment, i.e., the colors and frequencies of objects in jars and sample containers. The input-output pair coding is thus realistic, and it does not provide explicit information on the probabilities themselves. The output probabilities that we analyze are an emergent property of network learning.

Table 1. Schematic binary coding of a 4:1 vs. 1:4 frequency distribution.

| 4:1 | | 1:4 | |
|---|---|---|---|
| Input | Output | Input | Output |
| 1 | 1 | 2 | 1 |
| 1 | 1 | 2 | 0 |
| 1 | 1 | 2 | 0 |
| 1 | 1 | 2 | 0 |
| 1 | 0 | 2 | 0 |

An interesting feature of constructive learning in SDCC is that the interior of the network is constructed automatically, bypassing the need to hand-design a network. Recruiting of hidden units stops when it no longer improves performance, measured by stagnation in error reduction. Thus, each network is well suited to the complexity of the problem it is solving. A network's final computational power is sufficient to solve the training problem, and only as powerful as needed to minimize error, ensuring that it will not overtrain, at the expense of decreasing generalization.

In principle, a variety of other static neural-network learning algorithms could also learn probability distributions. However, it would be difficult to know in advance how to hand-design these alternate networks in terms of layers, connection patterns, and numbers of units, as these would vary with the complexity of the probability distributions being learned. This could presumably require considerable experimentation and tuning of networks for each unique probability distribution, essentially a human search of the network space in addition to an algorithmic search of the connection-weight space. In contrast, both searches are performed automatically and simultaneously in SDCC, thus elevating the psychological plausibility of NPLS. Instead of design expertise residing in human programmers, it resides within the algorithm. For similar reasons, constructive algorithms like SDCC make for more plausible computational models of learning probability distributions by biological agents. As detailed in the next section, NPLS can then induce a probability distribution on the learned mapping, affording probabilistic generativity.

It is worth noting that NPLS relies on a primitive notion of objects, in both learning and representation. The training and test patterns explicitly represent an object with three relevant properties: solidity, location, and color. All simulated objects are uniformly the same in our input example coding, except for being in a particular container, and their color, which varies across the contents of typically two containers. This is consistent with the view that a basic notion of *object* is a core property of early infant cognition (Spelke, 2000). We are agnostic about the origin of that knowledge.

**Sampling from a Learned Network**

An ordinary deterministic, feed-forward neural network can be converted into a probabilistic generative model by applying an MCMC algorithm to a trained neural network (Nobandegani & Shultz, 2017). Adopting this broad framework, NPLS induces a probability distribution $p(\mathbf{X}|\mathbf{Y})$ on the deterministic input-output mapping $f(\mathbf{X}; W^*)$ learned by an NPLS neural network, and then uses an MCMC process to sample from the induced distribution.

The induced distribution is given by:

$$p(\mathbf{X}|\mathbf{Y} = Y) \propto exp(-\beta||Y - f(\mathbf{X}; W^*)||_2^2) \qquad \text{Equation 4}$$

where $||\cdot||_2$ is the $l_2$-norm, $W^*$ the set of weights for a network after training, and $\beta$ a damping factor. For an input instance $\boldsymbol{X} = X$ belonging to the desired class $Y$, the output of the network $f(X; W^*)$ is expected to be close to $Y$ in the $l_2$-norm sense. Equation 4 adjusts the probability of input instance $X$ to be inversely proportional to the base-$e$ exponentiation of the $l_2$ distance.

MCMC processes are a class of algorithms that stochastically generate samples from a probability distribution of interest (aka the target distribution) and are used in modeling a range of human behaviors, e.g., cognitive biases and fallacies (Dasgupta, Schulz, & Gershman, 2017; Lieder, Griffiths, Huys, & Goodman, 2018), longstanding paradoxes in decision-making (Nobandegani & Shultz, 2020), and human generative abilities (Nobandegani & Shultz, 2017, 2018).

As we show in recent work (Nobandegani & Shultz, 2017), MCMC stochastically generates samples from the target distribution given in Equation 4, allowing NPLS to imagine new exemplars from a category of interest. Using an extension of the target distribution given in Equation 4, NPLS can imagine new exemplars from a category, under a wide range of hard and soft constraints (Nobandegani & Shultz, 2018). To accomplish this, NPLS incorporates these constraints into the target distribution in Equation 4 as a set of inductive biases.

MCMC processes start with some arbitrary initial sample and continue by stochastically generating samples ad infinitum. The sequence of samples generated by MCMC is guaranteed to asymptotically converge to the target distribution. That is, a sample generated later in the sequence comes from a distribution that more closely matches the target distribution. Hence, to have good-quality samples, MCMC processes are run to generate a long sequence of samples, with the final sample being considered as the best quality sample (in the sense that its distribution is closest to the target distribution) and the samples generated at the beginning of the sequence being considered as burn-in samples to be discarded. In the context of MCMC, constraints on cognitive resources can be effectively modeled by bounding the number of burn-in samples, with smaller numbers of burn-in samples corresponding to more severe cognitive constraints (Dasgupta et al., 2017; Lieder et al., 2018).

Our NPLS system can handle any MCMC process, including Metropolis-Adjusted Langevin, a gradient-based MCMC process, which can be implemented in a biologically-plausible manner (Moreno-Bote, Knill, & Pouget, 2011; Savin & Denève, 2014). In our simulations, we employ the Metropolis-Hastings (MH) MCMC process, used in modeling several aspects of cognition, including cognitive biases and decision-making paradoxes (Dasgupta et al., 2017; Lieder et al., 2018; Nobandegani & Shultz, 2020).

Like other MCMC processes, MH-MCMC starts with an arbitrary initial sample, and continues by stochastically generating samples ad infinitum, with a guarantee that the distribution of samples asymptotically converges to the target distribution. In our simulations, the target distribution is set to the one given in Equation 4. At each sample-generation step, MH-MCMC proposes a random sample (using a proposal distribution), compares the likelihood of this newly proposed sample to the likelihood of the sample generated in the previous step (where likelihoods are evaluated using the target distribution given in Equation 4), and shows a marked tendency to pick the sample with the highest likelihood. This tendency allows drawing both likely and unlikely samples (where likelihoods are evaluated using the target distribution), but generation frequency is approximately proportional to likelihood. This approximation improves as more samples are generated. The sampling results reported here use the classic MH-MCMC algorithm, with a uniform proposal distribution and $\beta = 2$ in Equation 4.

This sampling scheme, operating on learned neural-network connection weights, allows for bidirectional probabilistic reasoning in NPLS. A neural network learns probability distributions from realistic perceptual examples in a standard forward direction. The MH-MCMC sampling mechanism then reasons in the opposite direction, using the learned connection weights to accurately generate samples of the sort used in learning. This is a single, unified system in which accurate sampling benefits from learning, and learning can further benefit from actively querying for probabilistically generated samples (Yu, Nobandegani, & Shultz, 2019). There are more details on MH-MCMC in section SM1 of Supplemental Material.

## Simulations 1-2: Generalizing Between Samples and Populations

An early study of probability learning showed that infants could generalize from samples to populations and also from populations to samples (Xu & Garcia, 2008). There were two sets of experiments. The first set tested whether 8-month-olds could use the information in a sample to make inferences about a larger population. The second set tested the converse direction, asking if infants can use information about a population to predict a sample drawn from the population. In both sets of experiments, infants were first given a few ping-pong balls to play with for a few seconds, followed by four familiarization trials, and subsequent test trials. On each familiarization trial, a large box was brought onto the stage. Then, the experimenter opened the front panel of the box and drew the infant's attention to the box's contents. The box contained either mostly red ping-pong balls and a few white ping-pong balls, or mostly white ping-pong balls and a few red ping-pong balls. These two displays were shown to the infants on alternating trials to ensure that the infants were equally familiarized with each display. Then the test trials began.

For the first set of experiments testing sample-to-population generalization, the test trials were structured as follows. On each test trial, the same box was brought onto the stage, its contents unknown to the infants. The experimenter then shook the box and pulled out a ping-pong ball through the box's top opening. The experimenter then placed the ball into a transparent sample display container next to the large box. A total of five balls were drawn from the box, one at a time. In half of the test trials, a sample of 4 red balls and 1 white ball were drawn. In the other half of the test trials, a sample of 1 red and 4 white balls were drawn.

Then the experimenter opened the front panel of the box to reveal its contents, and the infant's looking time was recorded. Subsequently, the experimenter cleared the stage and started the next test trial until a total of 8 test trials were performed. The two samples (4:1 or 1:4 red) were shown to the infants on alternating test trials. To an infant who saw the mostly red outcome display when the box was opened, a 4:1 sample was more probable (hence expected), whereas a 1:4 sample was less probable (hence unexpected). The converse was true for an infant who saw the mostly white display when the front panel of the box was opened. Looking time measurements confirmed that 8-month-olds were more surprised on seeing a box with a majority color (e.g., red) producing a sample with the opposite minority color (e.g., white), indicating generalization from sample to population.

For the second set of experiments testing infants' population-to-sample generalization, the test trials were structured as follows. In each test trail, the experimenter brought the box onto the stage, opened the front panel of the box, and let the infant look at its contents for 5 s. Again, the box contained either mostly red or mostly white ping-pong balls. The experimenter then closed the front panel of the box, shook the box, closed her eyes, and reached into the box's top opening. Five ping-pong balls were drawn from the box, one at a time, and placed into the transparent sample display container next to the large box. On alternate test trials, the experimenter pulled out either 4 red balls and 1 white ball or 1 red and 4 white balls. Upon placing the fifth ball in the sample display container, the infant's looking time began to be recorded. The experimenter then cleared the stage and started the next test trial until a total of four test trials were performed. To an infant who saw the mostly red outcome display when the front panel of the box was opened, the 4:1 red sample was more probable (hence expected), whereas the 1:4 sample was less probable (hence unexpected) to have been randomly drawn from the box. The converse was true for an infant who saw the mostly white outcome display when the front panel of the box was opened. Looking time measurements confirmed that 8-month-olds were more surprised on seeing an unexpected sample being randomly drawn from the box, indicating that 8-month-olds generalized from populations to samples.

We train 20 NPLS networks in each of the two inference directions on samples illustrating the 4:1 and 1:4 color ratios of the samples drawn in the two sets of infant experiments (Xu & Garcia, 2008). Pictures of the test boxes (their Figures 1 and 3) suggest that about 6 minority-colored balls would be visible to the infant in an uncovered box. Consequently, we represent each visible box with a total of 6 replications of a 4:1 or 1:4 sample pattern. Training or testing on the total numbers of balls in each box may not be as realistic because many balls were not visible. However, increasing or decreasing the number of samples in a box representation does not significantly change our results. Additional samples representing a box are created by reproducing a pattern for a realistic number of times (6 in this case) to mimic how the box appears to an infant. In the sample-to-population condition, the network is trained on a single five-ball sample and then tested on 6 five-ball samples. In the population-to-sample condition, the network is trained on 6 five-ball samples and then tested on a single five-ball sample. The 6 samples of five balls represent the box contents, while the single sample of five balls represents the sample drawn from the box by the experimenter. Learning-cycle threshold is set to .01, and learning-cycle patience is set to 2.

A useful parameter in NPLS is score-threshold (ST), which we set to .5 in these simulations. Technically, ST is the maximum distance from target training values (in this case 0 or 1) considered to be correct. The default value of ST in NPLS is .4, providing a region of uncertainty around the .5 midpoint of the asymmetric sigmoid activation function. NPLS networks are run in learning-cessation mode to ensure that they quit learning when no further progress is being made in error reduction.

Simulation results are shown in Figure 1. Networks learn the probability distributions accurately and register more surprise at unexpected than expected outcomes. In the sample-to-population simulation, total error on all 6 population samples is divided by 6 to control for the number of samples being tested. Paired-sample $t$ tests show that output activations are higher for the 4:1 ratio than the 1:4 ratio in the sample-to-population simulation, $t(19) = 12185$, $p < .001$, $M = .600$, 95% CI [.600, .600] and the population-to-sample simulation, $t(19) = 1353$, $p < .001$, $M = .600$, 95% CI [.600, .601]. Also, network error is greater for unexpected than expected outcomes in the former simulation, $t(19) = 12111$, $p < .001$, $M = -3.600$, 95% CI [-3.600, -3.599] and latter simulation, $t(19) = 1341$, $p < .001$, $M = -3.600$, 95% CI [-3.605, -3.594]. All $p$ values in this article represent 2-tailed comparisons.

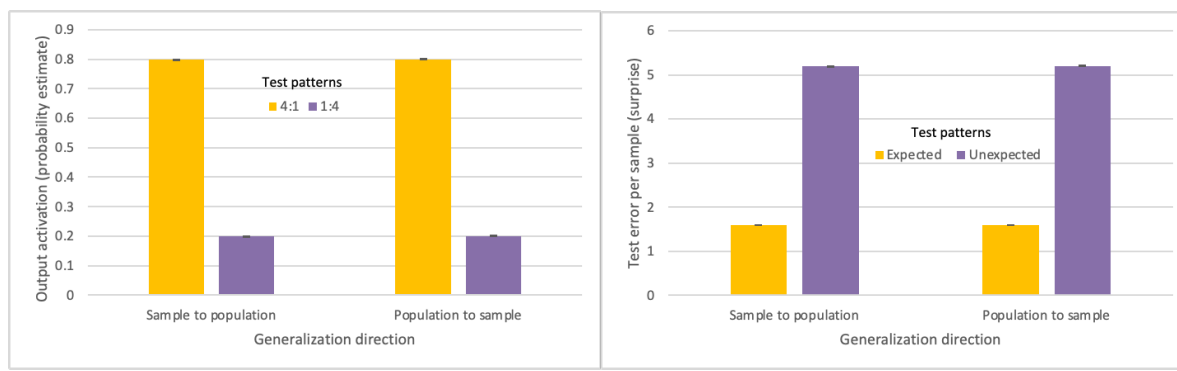

Figure 1. Mean output-activation probability estimates (left) and surprise at expected vs. unexpected outcomes (right) for generalizations in each direction. Error bars indicate standard deviations (SDs).

Network results are stronger than those reported for infants, reflecting strong learning even with an ST substantially greater than the default value of .4. Raising ST even higher than .5 would more closely approximate infant performance. The finding that both infants and networks generalize well in both directions supports the idea that learning focuses on the probabilities of events and not their frequencies, because the learning frequencies are 6 times greater for populations than for samples.

NPLS closely simulates the learning of ground-truth probability distributions and the patterns of infant surprise at expected and unexpected outcomes, all as emergent properties of neural-network learning. Other researchers (Teglas et al., 2011) simulated these infant experiments (Xu & Garcia, 2008) with a symbolic Bayesian model of intuitive physics but did not address the issue of how the probability distributions are learned.

In Supplemental Material section SM2, we include a follow-up simulation with the 4:1 and 1:4 ratios to study the effects of varying the patience parameter for learning cycles. Increasing learning-cycle patience to 8 allows more learning time, and recruitment of unnecessary hidden units, but does not improve learning accuracy or enhance differential surprise to unexpected events. It therefore seems reasonable to keep learning-cycle patience at a value of 2, which we do for all simulations in the present work.

### Simulation 3: Emergence of Probabilistic Reasoning in Young Infants

Another infant experiment focused on the development of the ability to learn and use probability distributions (Denison et al., 2013). In that experiment, 4.5- and 6-month-olds were shown two boxes, one containing a ratio of 1 pink to 4 yellow balls, the other containing the opposite ratio. The experimenter drew from, say, the mostly yellow box, gradually removing a sample of either 1 pink and 4 yellow balls (expected) or 4 pink and 1 yellow balls (unexpected) on alternating trials. Only the older infants looked longer at an unexpected, improbable sample than at an expected, probable sample.

Because depth of learning, manipulated by the score-threshold (ST) parameter in SDCC, has been shown to capture many developmental phenomena (Shultz, 2011, 2012), we set ST to 0.5 to represent the apparent deeper learning of the older infants and the higher value of 0.63 to represent the apparent shallower learning of the younger infants. Technically, ST is the maximum distance from target training values (in this case 0 or 1) considered to be correct.

In both conditions, NPLS runs in learning-cessation mode, which ensures quitting when no further progress is being made in reducing network error. We train 20 NPLS networks in each condition on 10 samples illustrating the 1:4 or 4:1 color ratios of the boxes from Denison et al. (2013). After training, the networks are tested on 5 samples representing either expected or unexpected outcomes. Figure 2 left shows that these probability distributions are accurate only with the deeper learning characteristic of the older infants. Paired-sample *t*-tests show a large difference for an ST of .5, $t(19) = 2759$, $p < .001$, $M = .600$, 95% CI [.600, .601] but not for an

ST of .63, *t*(19) = .933, *p* = .362, *M* = .034, CI [-.042, .110]. NPLS recruits 3 hidden units with a

ST of 0.5 and 0 hidden units with an ST of 0.63.

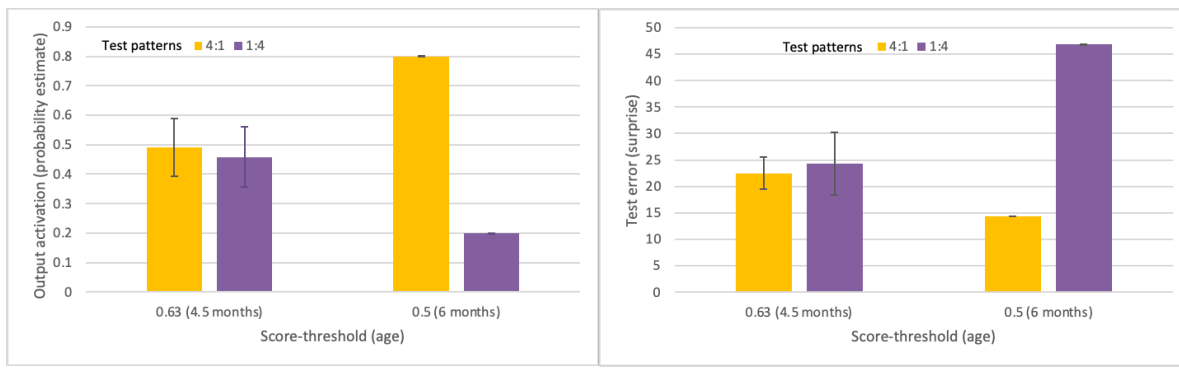

Figure 2. Probability estimates (left) and surprise to expected and unexpected samples (right).

Error bars indicate SDs.

Error on test patterns represents surprise at seeing an unexpected event, in this case an improbable sample of 5 balls. Figure 2 right shows that this surprise is noted only by networks which successfully learn the probability distribution, with ST of 0.5. As with output activations,

paired-sample *t*-tests show a large difference in network error for an ST of .5, *t*(19) = 2749, *p* <

.001, *M* = -32.415, 95% CI [-32.440, -32.391], but not for an ST of .63, *t*(19) = .93, *p* = .36, *M* =

-1.825, 95% CI [-5.919, 2.268].

## Simulation 4: Using Probabilistic Knowledge

Another experiment investigated the crawling patterns of 12- to 14-month-olds, providing

the first evidence for infants' ability to appropriately act on probabilistic expectations (Denison

& Xu, 2010). There was first a preference trial to determine whether infants preferred pink or black fake lollipops. Then there was a test trial wherein infants saw two jars with fake lollipops, one containing mostly pink lollipops and another containing mostly black lollipops. The pink ratios were 4:1 and 1:4, respectively. The experimenter removed one occluded lollipop from each jar and placed them in two separate opaque cups. The main finding was that 78% of infants searched in the cup that contained a lollipop from the jar with a higher proportion of their preferred color. This was very close to the ground-truth probability of .8 and reliably greater than chance, $t(31) = 3.79$, $p < .001$. The authors argued that this search task appeared to be a more sensitive measure than the standard, looking-time task that was designed to measure surprise.

Such infant selection patterns can be mathematically characterized as a form of sampling from the underlying 4:1 and 1:4 probability distribution. The infant would presumably mentally draw a sample, cued by the high probability of a favored object, yielding the identity of the jar with a higher proportion of that object. This sampling could then guide physical search towards the cup that was supplied by that jar.

Learning of this probability distribution is well-modeled by the population-to-sample experiment of Xu and Garcia (2008) in our first simulation, as both experiments used an identical 4:1 and 1:4 probability distribution and generalization from population to sample. But here, instead of surprise, we use the MH-MCMC process model to simulate sampling from this learned distribution. Example generation results are presented in Figure 3 averaged over 20 networks each generating $10^5$ samples. A paired-sample $t$ test confirms that NPLS mean sample probability is larger for favorable than for unfavorable cups, $t(19) = 846$, $p < .001$, $M = .536$, 95% CI [.535, .537].

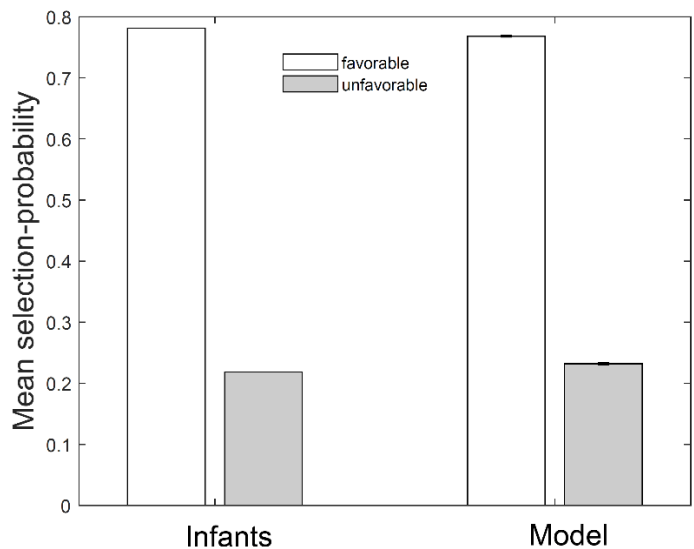

Figure 3. Probability of selecting the favorable and unfavorable cups by infants (Denison & Xu, 2010) (left), and the MH-MCMC process model (right), averaged over 20 NPLS networks each generating $10^5$ samples. Error bars indicate SDs.

To our knowledge, this provides the first computational explanation of these empirical findings. In doing so, our model also offers a rational process-level explanation for the between-subject variability observed in the infants' data (Denison & Xu, 2010), by showing that it follows from an MCMC process model with asymptotic convergence guarantees. Our model is consistent with a substantial body of work in developmental psychology providing evidence for the sampling abilities of children as young as four years (Bonawitz, Denison, Gopnik, & Griffiths, 2014; Denison, Bonawitz, Gopnik, & Griffiths, 2010).

### Simulations 5-8: Unconfounding Probability and Frequency

All of the foregoing simulations concern infant experiments in which frequencies of object types were thoroughly confounded with probabilities. That is, the most likely container for drawing a particular color also contained more items of that color. This raises the issue of whether infants might be using raw item frequencies rather than computing probabilities.

Thus, four subsequent, carefully-designed experiments (Denison & Xu, 2014) eliminated that confound, documenting that 10-12-month-old infants clearly use probabilities. In the unconfounded experiments (Denison & Xu, 2014), infants saw two colors of objects in live displays (Figure 4). Infants crawled to one of the two objects, revealing a preference for one or the other color. Then infants saw two jars containing different proportions of these object colors. The jars were then covered, and one object was randomly removed from each jar and hidden in a separate cup, without revealing its type. In each of four unconfounded experiments, infant searches reflected ground-truth probabilities and not raw frequencies.

The jar ratios of preferred to unpreferred objects are shown in rows 1 and 2 of Table 2. Ground-truth probabilities for each jar in each experiment are shown in rows 3 and 4. In Experiment 1, there was an equal number of preferred objects in each jar population, but differing probabilities of obtaining a preferred object across jars. Experiment 2 pitted probabilities against frequencies, as the infants' preferred object type was more numerous but less probable. Experiment 3 disconfirmed the possibility that infants used a different heuristic, raw frequencies of unpreferred objects rather than proportions of preferred objects. Finally, Experiment 4 challenged infants to distinguish a more subtle probability difference: .8 vs. .6.

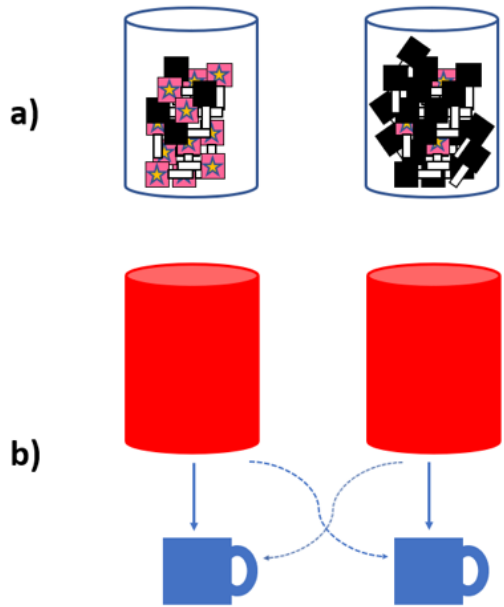

Figure 4. Schematic representation of a test trial in the infant experiments (Denison & Xu, 2014). (a) Infants were first familiarized with two populations with varying ratios of preferred vs. unpreferred objects. (b) The experimenter then removed an object from each jar one at a time and placed it in a cup. One-half of the infants saw the items placed in adjacent cups, while the other half saw the experimenter put each object in the slightly more distant (cross-over) cup. Finally, infants were encouraged to crawl to either cup, thereby revealing their understanding of which container and cup provided the better chance of obtaining their preferred object. Adapted from (Denison & Xu, 2014).

Here, we simulate these four infant experiments with NPLS for learning and sampling. As before, 20 NPLS networks in each of the four featured experiments (Denison & Xu, 2014) are trained on event sequences, with an input unit arbitrarily coding for the container (1 or 2) and an output unit coding for presence or absence of an object type (1 or 0, respectively).

Table 2. Ground-truth and simulated probabilities for each of the four experiments, with ratio reminders, and paired-sample *t* test comparisons of simulation means.

| Measures and results | Experiment | | | |
|---|---|---|---|---|
| | 1 | 2 | 3 | 4 |
| 1. Favorable ratio | 12:4 | 16:4 | 8:14 | 60:15 |
| 2. Unfavorable ratio | 12:36 | 24:96 | 8:72 | 60:40 |
| 3. Ground-truth favorable probability[a] | .75 | .80 | .36 | .80 |
| 4. Ground-truth unfavorable probability[b] | .25 | .20 | .10 | .60 |
| 5. Network favorable output-activation | .74 | .80 | .37 | .79 |
| 6. Network unfavorable output-activation | .26 | .20 | .12 | .59 |
| 7. *t*[c] for output activations | 18 | 536 | 16 | 19 |
| 8. *p* < for output activations | .001 | .001 | .001 | .001 |
| 9. Ground-truth favorable sampling *p*[d] | .75 | .80 | .78 | .57 |
| 10. Ground-truth unfavorable sampling *p*[d] | .25 | .20 | .22 | .43 |
| 11. Favorable sampling probability | .72 | .77 | .68 | .56 |
| 12. Unfavorable sampling probability | .28 | .23 | .32 | .44 |
| 13. *t*[c] for sampling | 15 | 68 | 16 | 10 |
| 14. *p* < for sampling | .001 | .001 | .001 | .001 |

[a] Ground-truth favorable probability is calculated as preferred-object frequency divided by total object frequency.

[b] Ground-truth unfavorable probability is calculated as unpreferred-object frequency divided by

total object frequency.

[c] *t* tests compare means for network favorable to unfavorable output-activations or samples.

[d] Ground-truth sampling probabilities for the binary searches in Experiments 3 and 4 are normalized by dividing the favorable and unfavorable ground-truth probabilities by their respective sums, because probabilities of mutually exclusive and exhaustive events, in this case searches, must sum to 1.

Table 2 integrates our main simulation results for both learning (rows 5-8) and sampling (rows 9-14) across the four experiments. For each experiment, the mean network probability estimates closely match ground-truth probabilities, consistent with the hypothesis that the infants were computing relevant probabilities (Denison & Xu, 2014). The mean estimated probability for the favorable location is considerably higher than that for the unfavorable location in every simulated experiment, as confirmed by a paired-sample *t*-test. Confidence intervals for these mean differences are presented in Table 3. Mean network output activations correlate highly with ground-truth probabilities across the eight conditions of the four experiments, $r(6) = 1.0$, $p < .001$, showing good coverage of pattern of means across conditions.

Table 3. Mean output-activation difference and 95% CI for each of four simulated experiments.

| Experiment | Mean difference | 95% CI |
|---|---|---|
| 1 | .473 | [.417, .529] |
| 2 | .600 | [.598, .602] |
| 3 | .249 | [.217, .281] |
| 4 | .190 | [.169, .211] |

One can more easily visualize the match between ground-truth probabilities and network output activations in Figure 5, which also includes standard deviation bars around the simulation means. In each experiment, the size and direction of the location difference between favorable and unfavorable activations is apparent (gold vs. purple) and very close to ground-truth probabilities (white vs. black). This represents the probabilistic knowledge that enables the sampling that could guide infant crawling towards the more favorable location for their preferred object. In contrast, predictions based on relative frequencies of the preferred object would expect no difference in Experiments 1, 3, and 4 (where those frequencies are equal across the two sources) and a reversed difference for Experiment 2 (where the preferred object is less probable when it is more frequent).

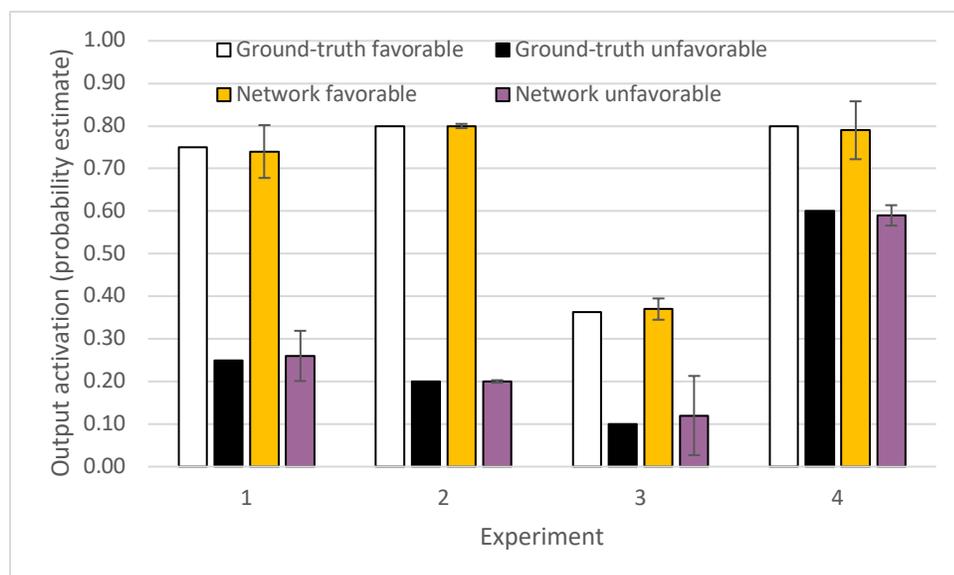

Figure 5. Ground-truth (white and black) and mean simulated probabilities (gold and purple; or light gray and dark gray in paper print) for each of four infant experiments. Error bars are standard deviations for 20 networks.

MH-MCMC applied to network weights accurately accounts for the infants' search patterns observed across the experiments (rows 9-14 of Table 2). We summarize our main results on infant crawling patterns in Figure 6, which shows a close match between infant crawling direction and mean samples drawn by the MH-MCMC algorithm operating on the connection weights learned by NPLS networks. These sample means estimate infants' probability of selecting the favorable vs. unfavorable jar, averaged over 1000 samples for each of 20 networks. In each experiment, the mean selection-probability for the favorable cup is higher than for the unfavorable cup (rows 13 and 14 of Table 2). Table 4 presents the mean selection-probability difference and 95% CI for each of these four experiments.

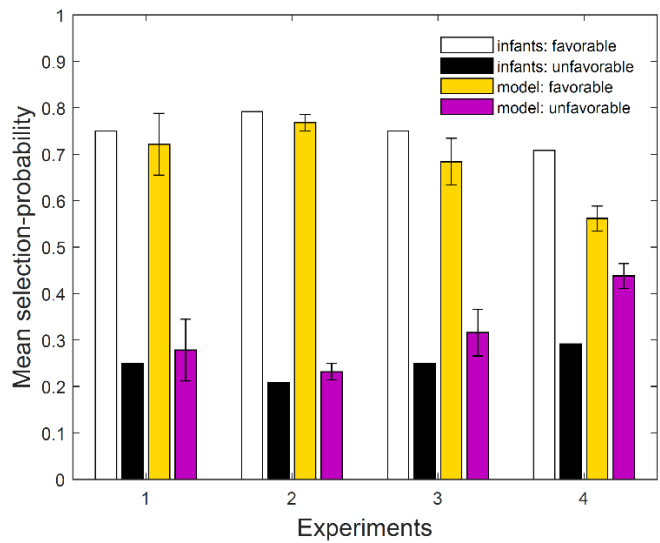

Figure 6. Probability of selecting the favorable and unfavorable cups by infants (white and black) (Denison & Xu, 2014) and the MH-MCMC process model (gold and purple; or light gray and dark gray in paper print), averaged over 20 networks each generating 1000 samples. Error bars indicate SDs.

MH-MCMC mean sample probabilities correlate highly with infant search probabilities across the eight conditions of the four experiments, $r(6) = .958$, $p < .001$. MH-MCMC mean sample probabilities also correlate highly with ground-truth probabilities across the eight conditions of the four experiments, $r(6) = .988$, $p < .001$. To compute the ground-truth sampling probabilities reported in Table 2 for Experiments 3 and 4 (corresponding to probability-matching behavior in binary searches) ground-truth probabilities are normalized, by dividing the favorable and unfavorable probabilities by their respective sums.

Table 4. Mean selection probability difference and 95% CI for each of each of four simulated experiments.

| Experiment | Mean difference | 95% CI |
|---|---|---|
| 1 | .433 | [.381, .505] |
| 2 | .536 | [.520, 553] |
| 3 | .368 | [.321, .415] |
| 4 | .124 | [.099, .149] |

In Supplemental Material section SM3-SM6, we present a deeper study of sampling in these simulations by imposing severe sampling limitations. The basic patterns hold even if very few samples are allowed.

### Simulation 9: Using Unpreferred Frequencies

Comparing rows 1 and 2 of Table 2, we notice a possible alternative hypothesis to explain the infant (Denison & Xu, 2014) and simulation data. In every experiment, the unpreferred frequency is substantially higher than the preferred frequency. It is thus possible that infants and/or networks are attending mainly to size of the unpreferred frequency and trying to avoid it in their search. There was an attempt to rule this out in the infant studies by substituting neutral objects for unpreferred objects in Experiments 3 and 4. There is a sense in which neutral objects are effectively unpreferred – they are different from the preferred object.

To provide a strong simulation test for ruling out this alternative hypothesis, we design and run a simulation experiment on this issue that could serve as a prediction for future infant experimentation. The simulation contrasts ratios in which either the unpreferred frequency (10:10 vs. 5:5) or the ratio (20:5 vs. 5:5) varies across conditions. The probability hypothesis predicts no difference in the former condition (ground-truth probability of .5), and a .8 vs. .5 (unnormalized) probability in favor of the more favorable ratio in the latter condition. In contrast, the unpreferred frequency hypothesis predicts the opposite: no difference in the latter condition and some unspecified degree of preference for the 5:5 source in the former condition, in order to better avoid unpreferred objects. Score-threshold is set to .53 in order to add more randomness.

Results are shown in Figure 7 for 20 NPLS networks in each of the two conditions. Independent samples t-tests indicate that only in the latter condition (right) is there a strong difference in favor of the source more likely to deliver a preferred object, $t(38) = 2954$, $p < .001$, $M = .300$, 95% CI [.300, .300]. There is no difference in the former condition (left), where ratio is constant and frequency of the unpreferred source varies, $t(38) = .192$, $p = .849$, $M = .001$, 95% CI [-.011, .014] . Again, this clearly supports the probability hypothesis over the frequency hypothesis for our model. Empirical researchers might want to see what happens in analogous infant experiments.

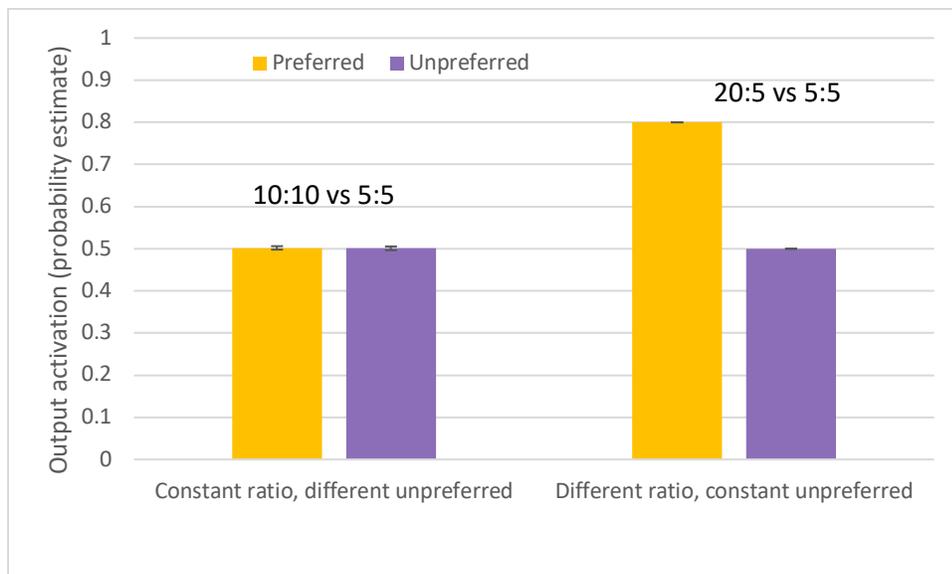

Figure 7. Mean estimated probabilities (and SDs) for a simulated experiment varying either unpreferred frequency (left) or ratio (right).

## Simulations 10-13: Reduced Ratios

The frequencies used in the infant experiments are rather large, making it unlikely that infants would process all of the available frequency information. It is more likely they would attend to only some of it. Indeed, some items visually obstructed by other items. We employ the same ratios used in the foregoing unconfounded four-experiment simulations, but here reduced to their smallest integer values. These reduced frequency ratios are shown in the last two rows of Table 5. Reduction loses the larger frequencies instrumental in unconfounding frequency and probability but enables examination of what happens when the ratios represent much smaller, and perhaps more realistic, samples presumably used by the infants in the brief time periods allowed in their experiments. We run 20 NPLS networks in a mixed-ANOVA design with the 4 experiments and 2 levels of ratio type (full vs. reduced) as between factors and 2 levels of favorability (favored vs. unfavored) as a repeated measure. The full-ratio data are from our original, unconfounded simulations (5-8).

Table 5. Original ratios and ground-truth probabilities, along with reduced ratios.

| Measure | Experiment | | | |
|---|---|---|---|---|
| | 1 | 2 | 3 | 4 |
| Favorable ratio | 12:4 | 16:4 | 8:14 | 60:15 |
| Unfavorable ratio | 12:36 | 24:96 | 8:72 | 60:40 |
| Ground-truth favorable probability | 0.75 | 0.8 | 0.36 | 0.8 |
| Ground-truth unfavorable probability | 0.25 | 0.2 | 0.1 | 0.6 |
| Reduced favorable ratio | 3:1 | 4:1 | 4:7 | 4:1 |
| Reduced unfavorable ratio | 1:3 | 1:4 | 1:9 | 3:2 |

The full ANOVA of output activation scores (probability estimates) reveals a large main effect of favorability, $F(1, 152) = 1106$, $p < .001$, $\eta_p^2 = .879$, reflecting higher preference for the cup supplied by the container with a higher probability of yielding the preferred object. There is also a smaller, 3-way interaction of favorability x ratio type x experiment, $F(1, 152) = 4.13$, $p = .008$, $\eta_p^2 = .075$, encouraging us to more deeply probe the favorability x ratio-type interaction separately for each experiment. The relevant means and SDs are plotted in Figure 8. For Experiments 1, 3, and 4, there are no main or interactive effects of ratio type. Experiment 2 is the exception, with a smallish favorability x ratio-type interaction, $F(1, 38) = 8.12$, $p = .007$, $\eta_p^2 = .176$. The Pearson correlation of mean output activation (probability estimates) with ground-truth probability values for reduced ratios, across the eight conditions of the four experiments, is $r(6) = .991$, $p < .001$. The overwhelming impression is that ratio reduction is just as effective for learning as using the full ratios is.

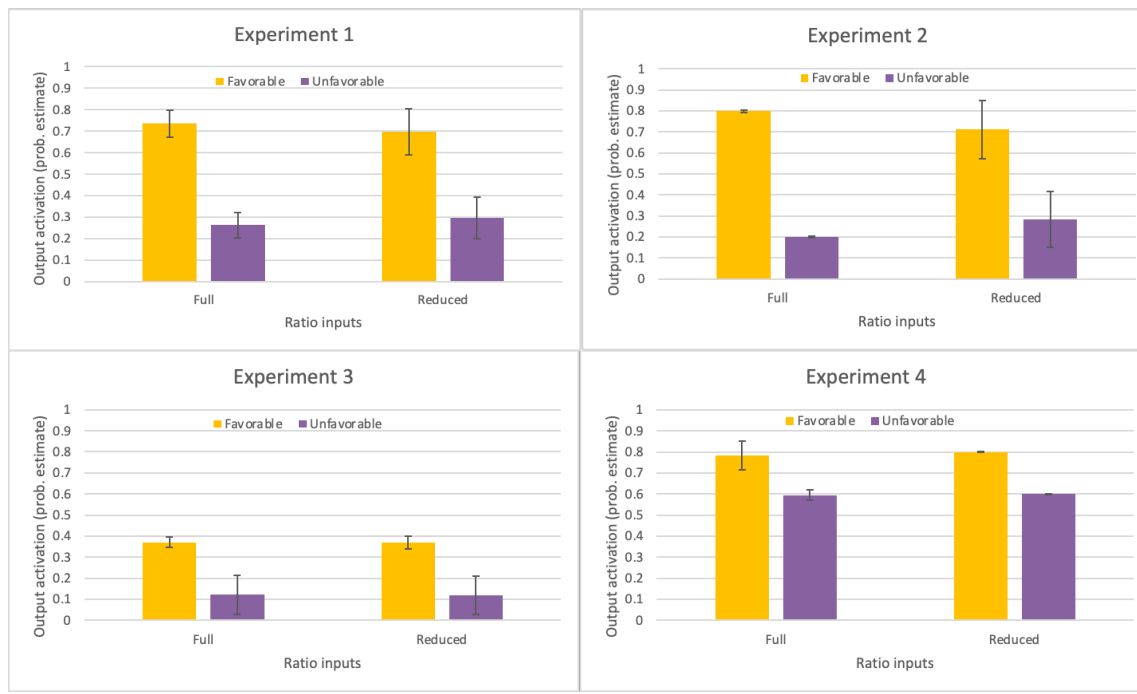

Figure 8. Mean output-activation probability estimates (and SDs) for 4 experiments with favorable or unfavorable sources x full or reduced ratios.

There is an effect of full vs. reduced ratios on learning time in two of the four experiments. In Experiments 1 and 2, learning takes longer for full than for reduced ratios, $t(38)$ = 3.4, $p$ = .002 for Experiment 1, and $t(38)$ = 4.1, $p$ < .001 for Experiment 2. Experiments 3 and 4 do not show this effect.

We conclude that small, representative samples are sufficient to produce basically similar learning results to the large-frequency samples used in the infant experiments (Denison & Xu, 2014) and our first simulations with full ratios. This accords well with our earlier conclusions that probabilities, not frequencies, are critical to explaining the infant data. Learning probability distributions from small, representative samples is just as effective as learning from large populations of samples. Both infants and NPLS networks focus on the relevant probabilities, ignoring the attendant frequencies.

## Simulations 14-15: Learning More than Two Probabilities

Nearly all of the infant probability research is so far restricted to only two probabilities at a time. In this section, to challenge the model a bit further, we explore the ability of NPLS to learn more than two probabilities. This introduces non-linearities requiring additional hidden units.

### Learning three probabilities

As a prediction for possible infant research, we first try three probabilities, with ratios of 2:8, 5:5, and 3:7, using input values of 1, 2, and 3, respectively, with score-threshold set to the default SDCC value of .4. Mean output activations, representing final probability estimates, are presented in Figure 9 for 20 NPLS networks, along with SDs. As with two-probability problems, learning is highly accurate with very small variance.  Activation means and their 95% CIs are shown in Table 6, computed as 1-sample *t*-tests.

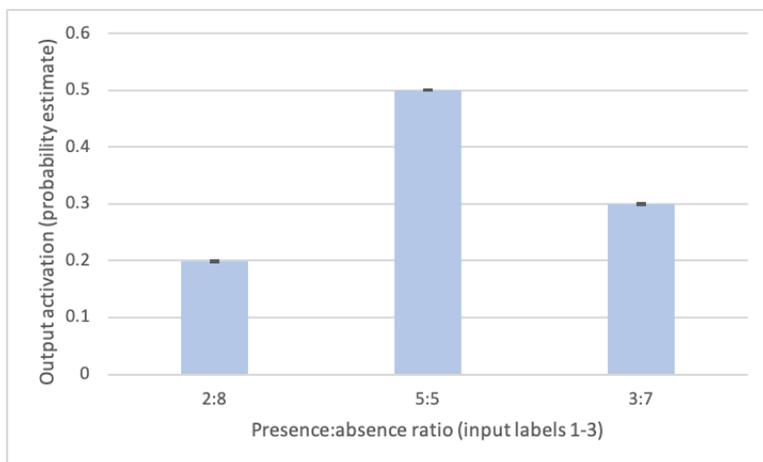

Figure 9. Mean probability estimates for 20 NPLS networks, and associated SDs, for each of three ratios learned simultaneously.

Table 6. 95% CI for the mean output-activation value at each ratio.

|       | Ratio |  |  |
|-------|----------|----------|----------|
|       | 2:8 | 5:5 | 3:7 |
| mean  | 0.199307 | 0.500593 | 0.299311 |
| lower | 0.198449 | 0.499952 | 0.298591 |
| upper | 0.200166 | 0.501234 | 0.300032 |

Learning does take longer with the extra, third probability, with a mean of 307 epochs, and SD of 29. These networks each recruit from 4-6 hidden units, with mean of 4.9 and SD of .45, installed on 1-3 layers. Without being able to recruit any hidden units on such probability problems with more than two probabilities, NPLS networks instead learn the mean of the three probabilities, in this case (.2 + .5 + .3) / 3 = .33.

**Learning a standard normal probability distribution**

Learning a continuous distribution of many probabilities (aka probability density function) is considerably more difficult than two or three probabilities, but NPLS seems up to that task as well. To make learning feasible, machine learning efforts often employ a discretized version of a target continuous distribution. Results for 20 NPLS networks learning the standard normal distribution (mean of 0 and SD of 1) are presented in Figure 10. For this demonstration, seven key sections of the distribution are included in the training set: the mean and 1, 2, and 3 SDs below and above the mean. There are 100 training patterns at each of these 7 ratios, consistently coded as event sequences as described in our Method section. Table 7 shows 95% confidence intervals for mean probability estimates, again using 1-sample *t*-tests.

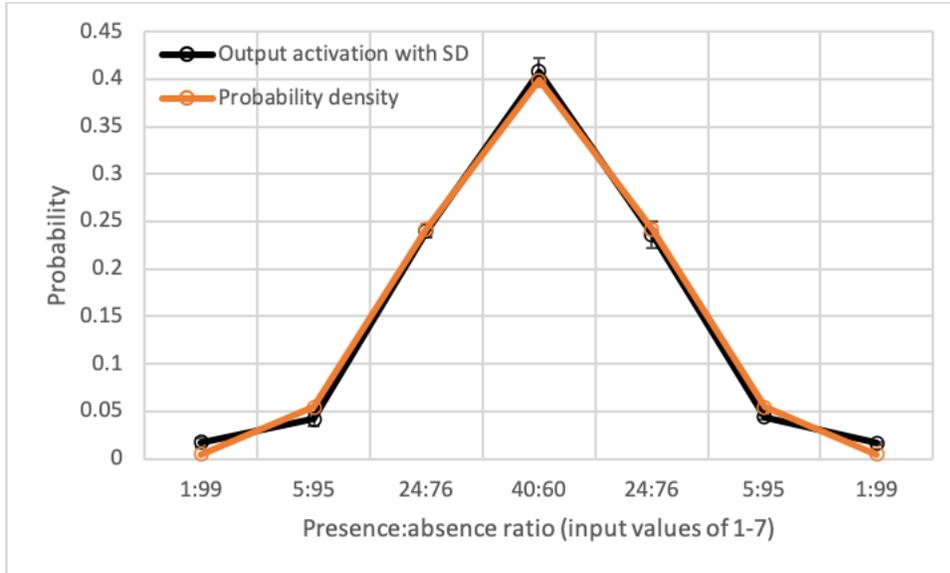

Figure 10. Mean probability estimates, and SDs, for 20 NPLS networks learning a discretized version of the standard normal probability distribution (black). Ground-truth probabilities for the probability density function are shown in orange. Training points, coded as event sequences as in our other simulations, are at the distribution mean and 1, 2, and 3 SDs below and above the mean.

Table 7. 95% CI for the output-activation mean at each input value.

| Input value | mean | lower | upper |
|---|---|---|---|
| 1 | 0.016956 | 0.014682 | 0.019230 |
| 2 | 0.041882 | 0.038682 | 0.045081 |
| 3 | 0.240302 | 0.237201 | 0.243402 |
| 4 | 0.041882 | 0.038682 | 0.045081 |
| 5 | 0.236429 | 0.229779 | 0.243080 |
| 6 | 0.043715 | 0.042593 | 0.044837 |
| 7 | 0.016246 | 0.015382 | 0.017111 |

Learning is accurate and is accomplished in a mean of 458 epochs (SD of 46), with a mean of 6.15 hidden units (SD of .59). There is a range of 5 to 7 hidden units placed on 2 to 4 layers. Mean output activations match the ground-truth probabilities with good accuracy, $r(5) =$ .998, $p < .001$. Again, if unable to recruit hidden units, NPLS networks learn only the mean of the probabilities, which in this problem is .14.

More of the curve in Figure 10 could be better approximated by finer discretization, requiring considerably more training patterns and probably longer learning. Agents might be better off representing such continuous curves by learning to estimate distribution parameters such as mean and variance.

It is clear that NPLS networks can learn considerably more than two probabilities in a single task. It would be interesting to employ a few more than two probabilities in infant experiments, as a further test of NPLS. Although we use the default score-threshold value of .4 in these demonstrations, raising score-threshold above .4 would show less accurate probability values with more variation, thus likely better approximating eventual infant data. Simulation results showing accurate sampling of examples and surprise at unexpected outcomes could also be added to such infant experiments.

## Discussion

In this work, we develop, instantiate, test, and analyze a model (NPLS) that provides a computationally sufficient explanation of how preverbal infants could learn and reason with probabilities without explicit, symbolic counting and dividing skills. From the deterministic encoding of the containers of colored objects seen by the infants, NPLS networks directly learn the relevant probabilities without explicitly counting frequencies and without explicitly dividing those counts by total frequencies. NPLS does this by autonomously building a network of the appropriate topology and adjusting its connection weights to reduce error, while predicting outcomes (here, presence or absence of colored objects) afforded by knowledge of stimulus sources (here, object containers). Other key features include using an asymmetric sigmoid activation function in the output unit (to keep outputs in the 0-1 range), learning cessation when error reduction stagnates (so that the network does not keep trying to learn probabilistic patterns ad infinitum), and pairing with an MCMC algorithm that uses network weights to probabilistically sample from the learned probability distribution. Here, NPLS is instantiated with MH-MCMC, but NPLS can handle any MCMC process.

The relevant probability distributions are not supplied to the networks as learning targets, but rather emerge naturally from reducing prediction error while learning event sequences. NPLS accurately represents infant visual scanning of the colored-object contents of two source containers. Our coding scheme for network learning realistically represents the results of these visual scans by pairing source-container identity with relative frequencies of the two colors. This is accomplished, not by explicitly counting these frequencies, but through binary coding of each object for presence or absence.

In Supplemental Material SM11, we present a mathematical proof that network output activations come to closely approximate the correct probabilities, providing mathematical insight into how and why NPLS networks can learn these simple binary probability distributions. According to our mathematical proof, binary coding of each object for presence and absence, together with optimally minimizing sum-of-squared error, logically implies the emergence of ground-truth probabilities at the network output. Importantly, the probabilities are not used as learning targets; they are instead emergent properties of a network of units that fire deterministically, depending on their input signals.

NPLS probability learning is based on minimizing sum-of-squared error (Method section, Equation 1). It is established in mathematical statistics that the maximum likelihood (ML) estimator of the mean of a normal distribution (which happens to be the optimal minimizer of sum-of-squared error) is the sample mean (Poor, 1994). NPLS probability learning is a special case of this principle, where observed stimuli are sequences of binary presences and absences (Table 1), resulting in the sample mean becoming the probability of object presence. To our knowledge, our work is the first to use this idea in modeling infant probability learning, where it is an effective way to learn probabilities without explicit symbolic counting and dividing.

The NPLS model provides a computationally sufficient model of a series of empirical experiments on probability learning and reasoning in infants. Following the sage advice of Newell (1994), we make no claims about NPLS's computational necessity. It is likely that some other current and future computational models could also convert stimulus ratio information into probability distributions, and thus successfully simulate the phenomena that we cover here.

Some of the likely contenders could be neural networks that build in probabilistic capacity by using units that fire stochastically, e.g., Boltzmann Machines (Ackley, Hinton, & Sejnowski, 1985), Restricted Boltzmann Machines (RBMs) (Hinton & Salakhutdinov, 2006), Deep Boltzmann Machines (Hinton, Osindero, & Teh, 2006), and Spike-and-slab RBMs (Courville, Bergstra, & Bengio, 2011). Such algorithms can learn a joint probability distribution over a set of input and latent variables. Because they are inherently stochastic, they have an important advantage over deterministic neural networks such as multilayer perceptrons, convolutional neural networks (Lecun, Bengio, & Hinton, 2015), and SDCC. However, due to a normalizing constant (aka partition function), learning in classic Boltzmann machines (Ackley et al., 1985) is computationally intractable and thus quite slow (Koller & Friedman, 2009;

Salakhutdinov, 2008). Because learning in NPLS does not involve estimating a partition function, NPLS learning is both quick and accurate, demonstrating that probabilities emerge in a fully deterministic system using units that fire deterministically.

There are also several proposals for how biological neural populations could encode probabilistic knowledge, e.g., distributional population codes (Ma, Beck, Latham, & Pouget, 2006; Zemel, Dayan, & Pouget, 1998), doubly distributional population codes (Sahani & Dayan, 2003), and efficient coding (Bhui & Gershman, 2018; Summerfield & Tsetsos, 2015). Future research should investigate whether and how these Boltzmann and population models could simulate the infant data considered here.

NPLS exhibits the computational sufficiency of techniques that infants likely possess, namely learning associations between stimuli (containers) and responses (color frequencies) by adjusting synaptic weights between neurons and recruiting additional neurons for as long as that helps to reduce prediction error. NPLS and infants also make inferences in the opposite direction – from the preferred color back to the identity of the source which is likely, in a random draw, to yield an object with that preferred color. Previous work with constructive neural networks using SDCC and its predecessor CC successfully simulated infant learning of a variety of deterministic tasks (Shultz, 2010; Shultz & Bale, 2001, 2006; Shultz & Cohen, 2004), underscoring a unified interpretation of a wide range of deterministic and stochastic phenomena.

In Supplemental Material, section Simulations SM7-10, we note that simple binary probabilities, like those common to infant experiments, can in principle be learned with only two weights. We confirm this phenomenon with a mathematical proof in section SM13, and note that SDCC networks closely approximate those ideal weights if allowed to learn for a sufficient time without being able to recruit any hidden units. However, if a few hidden units are recruited, learning is both faster and more accurate. Moreover, hidden units are required if the task involves more than two probabilities. A mathematical proof in SM12 shows that NPLS can learn any finite number of probabilities. Thus, standard NPLS provides a general and efficient compromise of accurate and fast learning across probability tasks of various complexity levels. Further, being a constructive learner, NPLS flexibly adapts its structure and computational power to the task at hand.

Moreover, neural networks with the most minimal structure (no hidden units) can lack robustness (Nobandegani, da Silva Castanheira, O'Donnell, & Shultz, 2019). In biological systems, robustness is the maintenance of functionality despite damage and loss. A common way to achieve robustness is redundancy (Félix & Barkoulas, 2015; Kitano, 2004). Recruiting hidden units, even if somewhat redundant, increases network robustness.

The recent, unconfounded infant experiments (Denison & Xu, 2014) and our simulations rule out an alternative explanation based on comparing only the raw frequencies of preferred objects. When the relevant probability patterns are not confounded with raw frequency patterns, both infants and NPLS focus accurately on probabilities and ignore frequencies.

By differentially including neutral items in the two target populations, Denison and Xu's (2014) Experiment 3 ruled out the alternative hypothesis that infants use quantities of unpreferred objects rather than proportions of preferred objects. We too rule out this alternative explanation with a rigorously controlled simulation in which NPLS focuses only on probability differences while ignoring unpreferred frequency differences. This constitutes a model prediction that could be tested in new infant experiments.

NPLS deserves further experimentation to better understand its computational properties, and its ability to simulate learning and use of probability distributions, not only in humans, but also in other species that have shown evidence of probability matching, including bees (Greggers & Menzel, 1993), fish (Behrend & Bitterman, 1961), turtles (Kirk & Bitterman, 1965), and apes (Eckert, Call, Hermes, Herrmann, & Rakoczy, 2018). Whether these other species use something like NPLS is something that we are starting to explore.

The fact that human infants and non-human species exhibit probability matching, rather than optimizing (by always choosing the more favorable option) is interesting and suggests a possible relation to the very active literature on the exploration vs. exploitation trade-off, a widely recognized issue in computer science, artificial intelligence, psychology, and behavioral economics (Agrawal & Goyal, 2012; Cohen, McClure, & Yu, 2007; Gershman, 2018; Wilson, Geana, White, Ludvig, & Cohen, 2014). NPLS does probability matching by sampling from learned probability distributions. The samples are generated probabilistically, and they approximate the NPLS-learned probability distribution on which the sampling is based. It seems likely that substantial probability learning could enable optimizing.

We find that learning is required in order to generate accurate samples. As explained in the Method section, NPLS integrates learning and sampling in a bidirectional process, forward from examples to probability distributions and then backwards from probability distributions to examples (Nobandegani & Shultz, 2017, 2018). In hundreds of networks simulating many empirical experiments, we have never seen an exception to the idea that successful learning is required for accurate constrained sampling. Whether and how infants might progress from probability-matching exploration to more optimal exploitation of their probabilistic knowledge should be further explored, in both empirical and modeling research.

Our work is potentially related to three other interesting literatures. There is an active literature on transition probabilities, particularly in language learning (Aslin, Saffran, &

Newport, 1998; Lany & Gómez, 2008; Saffran, Aslin, Johnson, & Newport, 1999). This is currently considered to be a distinctly different phenomenon (Denison & Xu, 2014), requiring a different kind of neural-network modeling in which auditory or visual graded chunks of information are gradually constructed and then recognized as they reappear in auditory or visual streams (Mareschal & French, 2017).

As well, there is an emerging literature on statistical summaries, aka *ensemble representation* (Alvarez, 2011). Visual cognition is enhanced when people (so far, only human adults) quickly summarize the statistical properties (e.g., average and variation) of a collection, affording an accurate and compact representation of the many items in the collection. This seems quite analogous to NPLS, in which a probability estimate compactly summarizes numerous presences and absences in event sequences.

NPLS, along with Boltzmann machines and neural population coding methods, encourages bridging between Bayesian and neural-network approaches and across different levels of analysis (Marr, 2010). Neural-network models operate at a lower, implementational level compared to the higher, computational level of Bayesian models. Each approach often partakes of an intermediate, algorithmic level. Some Bayesian researchers advocate bridging across these levels with MCMC sampling (Griffiths, Vul, & Sanborn, 2012). We agree that sampling is a key component, but we also find that some learning is often required to achieve accurate sampling. Additionally, it is important to understand how infants can learn various probabilities from rather brief exposures to large numbers of objects and event sequences.

Our sampling results in several simulations confirm that NPLS also performs well under realistic computational limitations, and are consistent with mounting evidence that children and adults alike often use only a few samples in probabilistic reasoning (Battaglia, Hamrick, &

Tenenbaum, 2013; Bonawitz, Denison, Griffiths, & Gopnik, 2014; Gershman, Horvitz, & Tenenbaum, 2015; Nobandegani & Shultz, 2020; Vul, Goodman, Griffiths, & Tenenbaum, 2014). Assuming that infants are indeed using sampling in probabilistic inference, it would be interesting to investigate how many samples infants actually process, perhaps with visual tracking technologies and varying exposure times. Our simulations suggest that, for the experiments modeled here, such variation in amount of sampling does not change the qualitative trend of infant results. This fits well with the idea that probability learning is indifferent to the magnitudes on which probabilities are based. How well this approach scales up to more complex abilities and probability distributions remains to be seen.

Our simulations offer several predictions that could perhaps guide further empirical research with infants: definitive evidence for use of probabilities over frequencies in critical experiments for which they make different predictions (Simulation 9), learning several probabilities simultaneously (Simulation 14), and the idea that ratio reduction does not change learning accuracy (Simulations 10-13).

Finally, it makes sense to assume that evolution has provided neural learning systems that can readily register and use a wide range of otherwise unanticipated probability distributions. NPLS is another step in the process of identifying and understanding such systems.

Supplemental Material for

A Computational Model of Infant Learning and Reasoning with Probabilities

Thomas R. Shultz and Ardavan S. Nobandegani

McGill University

This Supplemental Material contains further details on the Metropolis-Hastings MCMC algorithm, nine NPLS simulations, and three mathematical proofs.

- SM1: Details on the Metropolis-Hastings MCMC Algorithm

- Simulation SM2: Varying the Learning-cycle Patience Parameter

- Simulations SM3-SM6: Sampling Limitations

- Simulations SM7-SM10: Alternate Parameterizations

- SM11: Mathematical Proof Showing How and Why NPLS Learns Probabilities

- SM12: Mathematical Proof of Learning an Arbitrary Discrete Probability Distribution with Finite Support

- SM13: Mathematical Proof of Learning a 2-Probability Problem with Only Two Weights

**Metropolis-Hastings MCMC Algorithm**

The Metropolis-Hastings (MH) (Hastings, 1970; Metropolis, Rosenbluth, Rosenbluth, Teller, & Teller, 1953) process is a well-known Markov-chain Monte Carlo (MCMC) algorithm, which has been previously used in modeling several aspects of cognition, e.g., cognitive biases and fallacies (Dasgupta et al., 2017; Lieder et al., 2018), and decision-making paradoxes (Nobandegani & Shultz, 2020). Here is an outline of the MH algorithm:

**Input:** Target distribution $\pi(X)$, proposal distribution $q(X)$, number of samples $N$
**Output:** Samples $X_0, X_2, \dots, X_{N-1}$

1.  Pick $X_0$ arbitrarily
2.  **for** $i = 1$ **to** $N - 1$
3.  Sample $u \sim$ Uniform$[0,1]$
4.  Sample $X^* \sim q(X)$
5.  **if** $u < \min\{1, \frac{\pi(X^*)q(X_{i-1})}{\pi(X_{i-1})q(X^*)}\}$ **then**
6.      $X_i = X^*$
7.  **else**
8.      $X_i = X_{i-1}$
9.  **end if**
10. **end for**
11. **return** $X_0, X_2, \dots, X_{N-1}$

Here, $\pi(X)$ denotes the target distribution, $q(X)$ the proposal distribution, and $N$ the number of samples generated by MH. MH first picks an arbitrary initial sample $X_0$ (Line 1), and then continues by stochastically generating an arbitrary number of samples (Lines 2-11). At each $i$th sample-generation step, MH proposes a new sample $X^*$ randomly draw from the proposal distribution $q$ (Line 4). MH then compares the likelihood of this newly proposed sample $X^*$ to the likelihood of the sample generated in the previous step $X_{i-1}$ (where likelihoods are evaluated using the target distribution $\pi(X)$), and shows a marked tendency to pick the sample with the highest likelihood (Lines 5-9). MH follows this same procedure in every sample-generation step.

In our simulations, the target distribution $\pi(X)$ is set to:

$$\pi(X) = p(\mathbf{X}|\mathbf{Y} = Y) \propto exp(-\beta||Y - f(\mathbf{X}; W^*)||_2^2)$$

where $|| \cdot ||_2$ is the $l_2$-norm, $W^*$ the set of weights for a network after training, and $\beta$ is a damping factor. For an input instance $X = X$ belonging to the desired class $Y$, the output of the network $f(X; W^*)$ is expected to be close to $Y$ in the $l_2$-norm sense. This target distribution adjusts the probability of input instance $X$ to be inversely proportional to the base-$e$ exponentiation of the $l_2$ distance. For consistency, all the sampling results reported in this work use a uniform proposal distribution $q$, and $\beta = 2$ in the equation.

**Simulation SM2: Varying the Learning-cycle Patience Parameter**

In NPLS, a learning-cycle variable with its own threshold and patience parameters, is introduced to monitor learning progress and stop learning when progress stagnates. This is at the point when the probability distribution has just been well learned. Setting the threshold and patience parameters for learning-cycles is somewhat arbitrary and has not been well studied. The learning-threshold parameter is the proportion by which error must change in order to count as significant. The learning-patience parameter is the number of learning cycles the algorithm is willing to accept without quitting. For example, if patience is set to 2, learning will cease after 3 hidden unit recruitments, where success is defined by a significant error change. Typically, any significant error change will be a reduction of error. In the present work, we keep learning-cycle threshold at .01 (the same as the default value used for threshold in SDCC output-phases) and patience at 2 (quite a bit less than the default value of patience in SDCC output phases, which is 8). Threshold can be considered as sensitivity to changes in error, and it seems reasonable to keep it at a somewhat smallish value regardless of which time cycle (learning or output-phase) is being considered.

Here, we examine whether a learning-cycle patience of 2 can be further justified for our present simulations. One realistic consideration is that infants in the experiments we simulate have very limited exposure time with which to examine the contents of the relevant containers, perhaps only a few seconds. It is difficult to pinpoint this exactly because exposure time is typically only a part of a somewhat longer procedure.

In this simulation, we compare a learning-cycle patience of 2 with the default SDCC output-phase value of 8 epochs, using the 4:1 vs. 1:4 ratios featured in simulations 1-2. Twenty networks are run in each of the two ratio conditions, and we examine the effects of patience and ratio on four dependent variables: output activations, network error, output-phase epochs, and number of hidden units recruited. We use a score-threshold of .53 to introduce a bit of error into the results. As score-threshold increases, this introduces some fuzziness at the decision border. The training and test patterns each represent a single sample.

The means and SDs of output activations, representing network probability estimates, are shown in Figure SM1. A ratio x patience mixed repeated-measures ANOVA reveals only a main effect of ratio on output activations, $F(1, 38) = 276$, $p < .001$. As with Simulations 1-2, NPLS networks closely approximate the .8 vs. .2 ground-truth probabilities with their output-unit activation. The main effect of patience, $F(1, 38) = .983$, $p = .328$, and interaction effect of patience x ratio, $F(1, 38) = .211$, $p = .648$, clearly do not affect network probability estimates.

The same pattern is evident for network error, measuring surprise (Figure SM2). Networks simulate more surprise (measured by network error) for unexpected results than expected results, $F(1, 38) = 276$, $p < .001$, while neither patience, $F(1, 38) = .215$, $p = .645$, nor the interaction of patience and ratio, $F(1, 38) = .209$, $p = .65$, matters. Confidence intervals around the key means of probability estimates and surprise are presented in Table SM1.

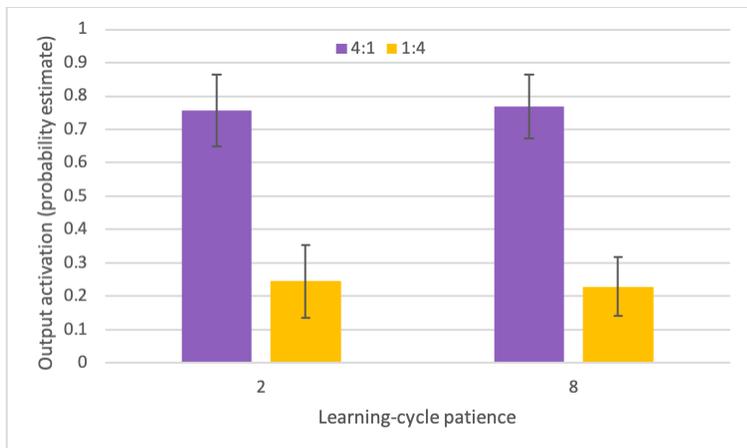

Figure SM1. Mean output-activation probability estimates as a function of ratio and learning-cycle patience. Error bars indicate SDs.

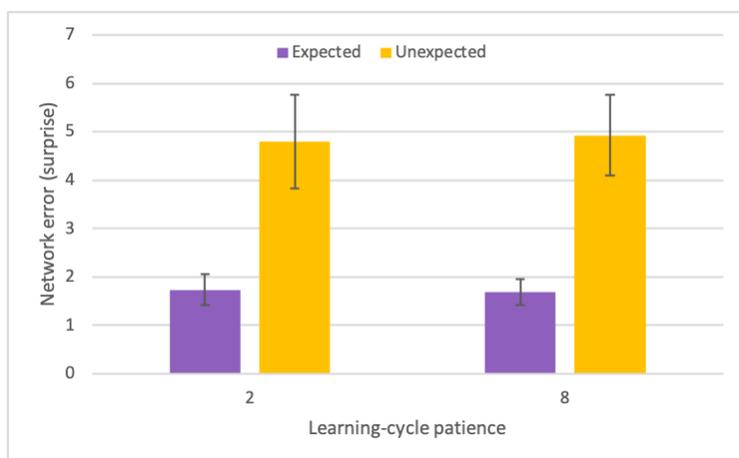

Figure SM2. Mean and SD of simulated surprise (network error) as a function of expectation and learning-cycle patience.

Table SM1. Means and 95% CIs for output-activation and network error at two levels of learning-cycle patience.

| Measure | Patience | Frequency ratio | Mean | Lower | Upper |
|---|---|---|---|---|---|
| Output activation | 8 | 4:1 | .769 | .723 | .815 |
| Output activation | 8 | 1:4 | .229 | .184 | .273 |
| Output activation | 2 | 4:1 | .756 | .710 | .802 |
| Output activation | 2 | 1:4 | .244 | .200 | .289 |
| Network error | 8 | 4:1 | 1.688 | 1.554 | 1.823 |
| Network error | 8 | 1:4 | 4.932 | 4.523 | 5.341 |
| Network error | 2 | 4:1 | 1.731 | 1.596 | 1.866 |
| Network error | 2 | 1:4 | 4.801 | 4.392 | 5.210 |

For learning time, a patience of 8 produces longer simulations than a patience of 2 (means of 466 vs. 165 epochs), $t(38) = 7.56$, $p < .001$, *M-difference* = 301, 95% CI [220, 382]. These extra epochs yield more recruitments of hidden units (means of 8.1 vs. 2.55), $t(38) = 8.33$, $p < .001$, *M-difference* = 5.55, 95% CI [4.201, 6.899].

In summary, increasing learning-cycle patience to 8 allows more learning time, and recruitment of unnecessary hidden units, but does not improve learning accuracy or enhance differential surprise to unexpected events. It therefore seems reasonable to keep learning-cycle patience at a value of 2. At this setting of 2 for learning-cycle patience, learning is both quick and accurate, as it is with the simulated infants. Hence, this setting is the default for the rest of the simulations.

## Simulations SM3-SM6: Sampling Limitations

Here, we investigate whether our sampling model can simulate infants' search patterns for the four-experiments article (Denison & Xu, 2014) under severe sampling limitations. Using the MCMC process to generate samples from the probability distributions learned by NPLS neural networks described in our original 4-experiment simulation, we simulate infant search patterns under relatively unconstrained cognitive limitations (operationalized by an MCMC burn-in period of 1000 samples, followed by a single sample indicating the model's prediction for the box selected by an infant, Figure SM3 left) and under severe cognitive limitations (operationalized by limiting the MCMC burn-in period to only one sample, followed by a single sample indicating the model's prediction for the box selected by an infant, Figure SM3 right). In MCMC sampling processes, the burn-in period refers to the set of initial inaccurate samples generated that are typically discarded, and hence, not used for probabilistic inference. As before, in each experiment, predicted mean selection-probability for the favorable cup is higher than for the unfavorable cup. Associated *t*-test results are displayed in Table SM2, and 95% CIs in Table SM3.

The MCMC mean sample probabilities also correlate highly with normalized ground-truth probabilities (corresponding to accurate probability-matching behavior) across the eight conditions of the four experiments both after 1000 burn-in samples, $r(6) = .989$, $p < .001$, and after one burn-in sample, $r(6) = .993$, $p < .001$. The MCMC mean sample probabilities also correlate highly with infant search probabilities across the eight conditions of the four experiments, whether after 1000 burn-in samples, $r(6) = .953$, $p < .001$, or after one burn-in sample, $r(6) = .966$, $p < .001$.

Thus, our MCMC sampling results are very accurate in matching both normalized ground-truth probabilities (corresponding to probability-matching behavior) and infant search patterns, whether a single sample is drawn after a lengthy or extremely limited burn-in period.

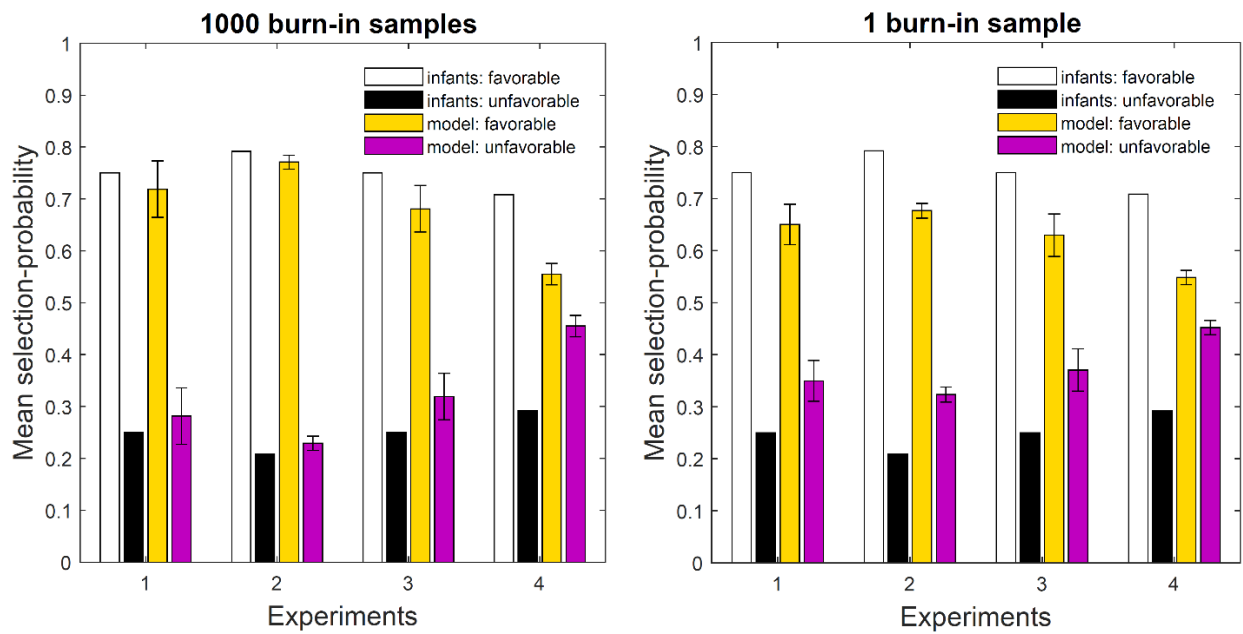

Figure SM3. Probability of selecting the favorable (white) and unfavorable (black) cups, as demonstrated by infants (Denison & Xu, 2014) and the MH-MCMC process model (gold and purple; or light gray and dark gray in paper print) averaged over 20 networks. MH-MCMC's burn-in period is set to 1000 burn-in samples (left), or only one burn-in sample (right). The sample generated immediately after the burn-in period indicates the model's prediction of the selected box. Error bars denote SDs.

Table SM2. Paired-sample *t*-test results comparing the gold vs. purple means in Figure SM3.

|  | 1000 burn-in samples | | 1 burn-in sample | |
|---|---|---|---|---|
| Experiment | *t*(19) | *p* < | *t*(19) | *p* < |
| 1 | 18 | .001 | 17 | .001 |
| 2 | 89 | .001 | 55 | .001 |
| 3 | 18 | .001 | 14 | .001 |
| 4 | 12 | .001 | 16 | .001 |

Table SM3. Mean selection-probability difference and 95% CI for each of four simulated experiments in Figure SM3.

|  | 1000 burn-in samples | | 1 burn-in sample | |
|---|---|---|---|---|
| Experiment | Mean difference | 95% CI | Mean difference | 95% CI |
| 1 | .437 | [.386, .488] | .301 | [.264, .337] |
| 2 | .542 | [.529, .554] | .354 | [.340, .367] |
| 3 | .362 | [.320, .404] | .259 | [.221, .297] |
| 4 | .110 | [.091, .129] | .096 | [.083, .109] |

**Simulations SM7-SM10: Alternate Parameterizations**

Learning a distribution of only two contrasting probabilities may be a fairly simple problem for neural networks. Indeed, it is in principle possible to learn these simple two-probability problems using neural networks without any hidden units. It can be done with just two connection weights to the output unit, one from the bias unit, which always has an input of 1, and another from the single input unit, which can encode the use of two different containers. See Supplemental Material section SM11 for a mathematical proof of this.

Here, we explore this idea empirically, comparing standard NPLS, which recruits as many hidden units as it needs, against three different parameterizations that learn with only two weights, as just described, recruiting no hidden units at all: one that runs for only 30 output-phase epochs, and two others that run long enough to recruit 3 hidden units as standard NPLS does, but do not actually recruit any hidden units. Essentially, the latter two models serve as a control for learning time, as compared to standard NPLS. All four parameterizations are learning the 4:1 and 1:4 ratios, the most commonly used problem in infant experiments. All but one of the four parameterizations run with the default score-threshold value of .4, while the exceptional parameterization runs with a much lower score-threshold of .001, in principle forcing a greater degree of learning accuracy. The three parameterizations that can train as long as they need to each have a learning-patience of 2. There are 20 networks in each of the four conditions.

Probability estimates for the four parameterizations are plotted in Figure SM4. A mixed ANOVA with 4 parameterizations as a between-run factor and 2 ratios as a repeated measure yields a variant x ratio interaction, $F(3, 76) = 21, p < .001, \eta_p^2 = .458$. This interaction reflects that fact that, although all four parameterizations learned the probability distribution, the one limited to 30 epochs is less accurate and more variable than the three parameterizations that learn longer.

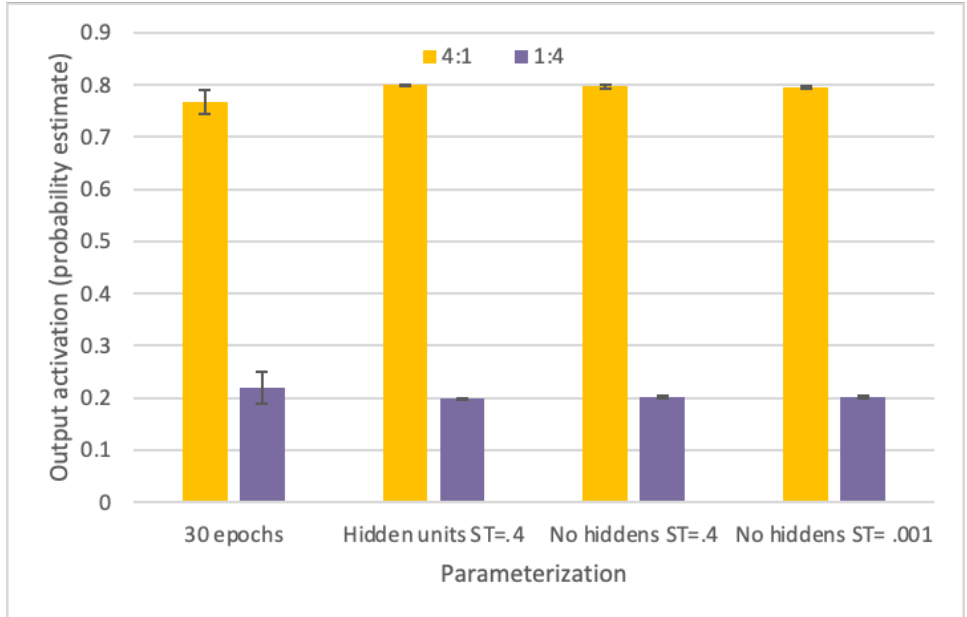

Figure SM4. Mean probability estimates (with SDs) for four different parameterizations.

This trend can be visualized more easily in Figure SM5, which presents mean activation difference scores, calculated by subtracting the 1:4 activation values from the 4:1 activation values. A one-way factorial ANOVA of these difference scores yields a main effect of NPLS variant, $F(1, 76) = 21$, $p < .001$, $\eta_p^2 = .458$. Both LSD and Tukey (HSD) multiple-comparison tests show that networks running for 30 epochs have smaller difference scores than each of the other three parameterizations, $p < .001$, which cluster in the same homogeneous subset. That is, the networks learning longer are more accurate than networks that learn very quickly.

Whether that extra learning involves only two output weights (from the bias and input units) or includes three recruited hidden units does not matter very much for accuracy. However, we can see in Figure SM6 that learning takes quite a bit longer without hidden units than with hidden units, $F(1, 57) = 28$, $p < .001$, $\eta_p^2 = .495$.

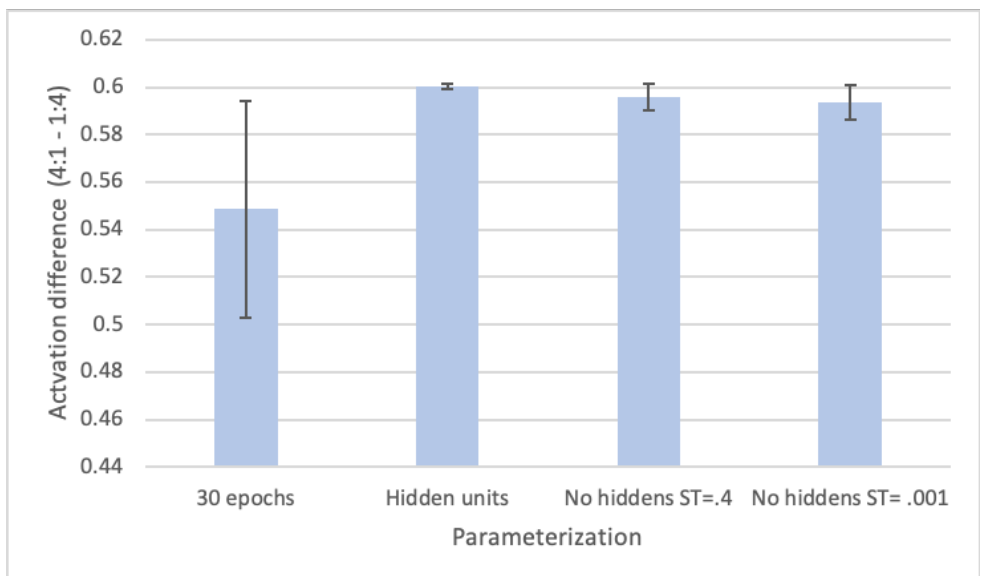

Figure SM5. Mean activation difference scores (4:1 – 1:4), with SDs, for four parameterizations.

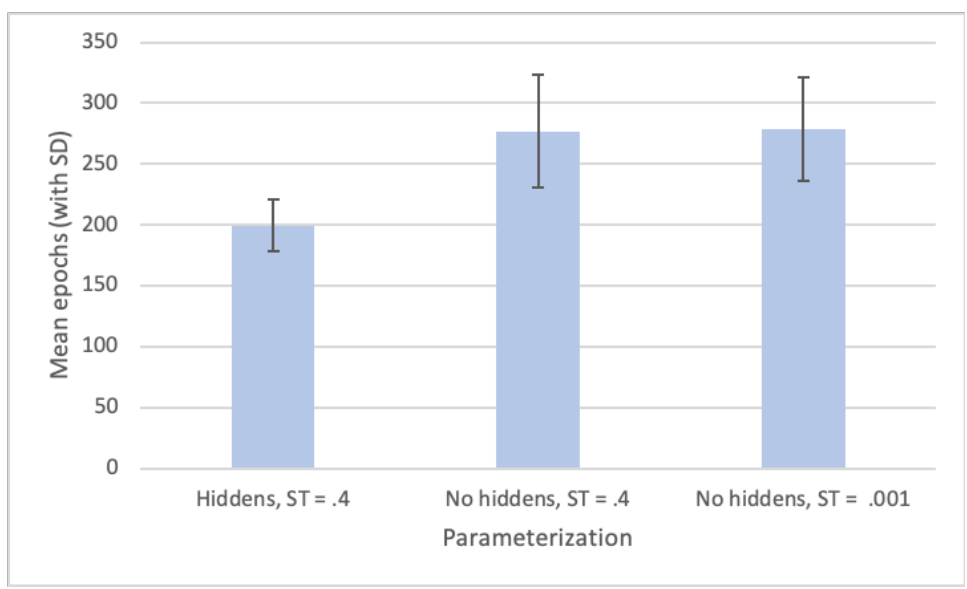

Figure SM6. Mean output epochs to learn (with SDs) for networks with vs. without hidden units.

Thus, even though all these four different parameterizations are computationally sufficient to learn a simple 2-probability problem sufficiently accurately, the standard parameterization in NPLS, allowing the recruitment of hidden units, provides the best combination of accuracy and speed. In contrast, the learning in networks without hidden units is either sloppier (networks learning for only 30 epochs) or slower (networks allowed additional time, whether seeking default or extreme accuracy). Moreover, as shown in section Simulations 14-15, NPLS networks definitely require hidden units to learn more than two probabilities.

Mean network connection weights from the bias and input units are compared in Table SM4 to those from the mathematical proof. The mathematically derived weights are most closely approximated by networks without hidden units that are allowed a long learning period (rows 3 and 4 in the table). In the case of networks recruiting hidden units, some of the functionality supporting accuracy is provided by those hidden units.

Table SM4. Mean network weights from the bias and input units and from our mathematical proof (SM12).

| Source | Weight from bias | Weight from input |
|---|---|---|
| 1. 30 epochs | 3.7253 | -2.4981 |
| 2. Hidden units | 3.7426 | -2.5641 |
| 3. No hiddens, ST = .4 | 4.1199 | -2.7489 |
| 4. No hiddens, ST = .001 | 4.1022 | -2.7376 |
| 5. Mathematical proof | 4.1589 | -2.7726 |

**SM11: Mathematical Derivation Showing How and Why NPLS Learns Probabilities**

The question of how and why our model simulates the infant results so well naturally arises. We attempt here to gain some mathematical insight into how our model simulates infant probability learning so accurately.

In our Method section, we noted that during NPLS learning, in what is called output phase, connection weights are altered so that network error is reduced. In Equation 1 of the main article, we saw that sum of squared error is computed as:

$$E = \sum_o \sum_p \left(A_{op} - T_{op}\right)^2$$

In this derivation, we show mathematically that our input-output coding scheme, together with learning cessation, and NPLS minimization of the sum-of-squared error in Equation 1, unfailingly leads to the emergence of network output values closely approximating ground-truth probabilities. The proof starts by separating network error into two natural parts: presence vs. absence of the event in question. Presented with a ratio $N_1 : N_0$, NPLS adjusts its topology and weights so that its output activation $x$ minimizes the sum-of-squared error $E$ for both present and absent examples:

$$E = \sum_{i=1}^{N_1} (x - 1)^2 + \sum_{j=1}^{N_0} (x - 0)^2 \quad (1)$$

The first term in Equation 1 represents the error attributable to examples where the output event is present (target output value of 1), while the second term represents examples where the output event is absent (target output value of 0). The learning algorithm jointly reduces both sources of error.

The optimal minimizer of $E$, $x^*$, has to satisfy the condition that the derivative of $E$, evaluated at $x = x^*$, is equal to 0:

$$\frac{d}{dx} E \rfloor_{x=x^*} = 0 \quad (2)$$

Error is at a minimum where that derivative is 0. Thus, we compute the derivative of Equation (1), set it to 0, and solve for $x^*$, after deleting - 0 in the second term:

$$2 \sum_{i=1}^{N_1} (x^* - 1) + 2 \sum_{j=1}^{N_0} x^* = 0 \qquad (3)$$

Factor out 2 and divide each side of the equation by 2.

$$\sum_{i=1}^{N_1} (x^* - 1) + \sum_{j=1}^{N_0} x^* = 0 \qquad (4)$$

Substitute equivalent multiplication for summation.

$$N_1 (x^* - 1) + N_0 x^* = 0 \qquad (5)$$

Multiply the first 2 terms.

$$N_1 x^* - N_1 + N_0 x^* = 0 \qquad (6)$$

Add $N_1$ to each side.

$$N_1 x^* + N_0 x^* = N_1 \qquad (7)$$

Factor out $x^*$.

$$x^* (N_1 + N_0) = N_1 \qquad (8)$$

To isolate $x^*$, divide both sides by $N_1 + N_0$.

$$x^* = N_1 / (N_1 + N_0) \qquad (9)$$

This derivation proves that the activation at the network's output will come to closely approximate the correct probability. It also shows how probabilities are learned by NPLS as an emergent property of neural-network operations, without the need for explicit counting and dividing of events. This provides mathematical insight into how and why NPLS networks can learn simple binary probability distributions. Critical features of NPLS include learning by error reduction, deterministic coding of event-sequence examples that collectively conveys a probabilistic pattern directly encoding the visual information presented to infants in each particular experiment, learning cessation when error reduction stagnates, and use of an output activation function in the 0-1 range. In Supplemental Material section SM12, we show that this proof can be generalized to tasks with more than two probabilities.

## SM12: Mathematical Proof of Learning an Arbitrary Discrete Probability Distribution with Finite Support

Here, we mathematically prove that our input-output coding scheme (see Method, main text), together with representative examples, sufficiently deep learning, learning cessation, and NPLS minimization of the sum-of-squared error in Equation 1, unfailingly leads to the emergence of network output values closely approximating ground-truth probabilities of any discrete probability distribution with finite support (i.e., a discrete probability distribution defined on a finite number of outcomes). Without loss of generality, we represent this target probability distribution by a ratio $N_1 : N_2 : N_3 : \ldots : N_n$, where $N_i$ denotes the frequency of the $i^{\text{th}}$ outcome and $n$ denotes an arbitrary, but finite, number of outcomes on which the probability distribution is defined.

Our proof is an extension of the proof presented in SM11 and follows the same line of reasoning. Again, we start by separating network error into two natural parts: presence vs. absence of the event in question, which is, for the purpose of this exposition, and without loss of generality, the first event. Presented with a ratio $N_1 : N_2 : N_3 : \ldots : N_n$, NPLS adjusts its topology and weights so that its output activation $x$ minimizes the sum-of-squared error $E$ for both present and absent examples:

$$E = \sum_{i=1}^{N_1} (x-1)^2 + \sum_{j=1}^{N_2 + \cdots + N_n} (x-0)^2 \qquad (1)$$

The first term in Equation (1) represents the error attributable to examples where the output event is present (target output value of 1), while the second term represents examples where the output event is absent (target output value of 0). The learning algorithm jointly reduces both sources of error.

The optimal minimizer of $E$, $x^*$, has to satisfy the condition that the derivative of $E$, evaluated at $x = x^*$, is equal to 0:

$$\frac{d}{dx}E]_{x=x^*} = 0 \qquad (2)$$

Error is at a minimum where that derivative is 0. Thus, we compute the derivative of Equation (1), set it to 0, and solve for $x^*$, after deleting - 0 in the second term:

$$2\sum_{i=1}^{N_1}(x^* - 1) + 2\sum_{j=1}^{N_2 + \cdots + N_n} x^* = 0 \qquad (3)$$

Factor out 2 and divide each side of the equation by 2.

$$\sum_{i=1}^{N_1}(x^* - 1) + \sum_{j=1}^{N_2 + \cdots + N_n} x^* = 0 \qquad (4)$$

Substitute equivalent multiplication for summation.

$$N_1(x^* - 1) + (N_2 + \cdots + N_n)x^* = 0 \qquad (5)$$

which simplifies to

$$(N_1 + N_2 + \cdots + N_n)x^* = N_1 \qquad (6)$$

Solving for $x^*$, we get

$$x^* = \frac{N_1}{N_1 + N_2 + \cdots + N_n} \qquad (7)$$

This derivation extends the proof presented in SM11 by showing that the activations at the network's output will come to closely approximate the correct probabilities of any discrete probability distribution with finite support. It also shows how probabilities are learned by NPLS as an emergent property of neural-network operations, without the need for explicit counting and dividing of events. This provides mathematical insight into how and why NPLS networks can learn any finite-support discrete probability distribution.

## SM13: Mathematical Proof of Learning a 2-Probability Problem with Only Two Weights

Here, we mathematically prove that both a simple, linear, two-unit neural network (with one input, one output, and zero hidden units) and also a sigmoidal, two-unit neural network can, in principle, learn any possibly unnormalized binary distribution.

Let us denote a generic binary distribution by $(1, p; 2, q)$ with $p, q > 0$, meaning that input 1 signals an event with probability $p$ and input 2 does so with probability $q$. The output $y$ of a linear, two-unit neural network is given by $y = wx + b$, where $w, x, b$ denote the input weight, the input, and the bias of the network, respectively. According to our earlier mathematical analysis, the minimizer of the sum-of-squared error is the ground-truth probability. That is, the sum-of-squared-error is minimized when, for each input, the output of the neural network is the corresponding ground-truth probability. Accordingly, to demonstrate that a two-unit neural network can, in principle, learn any possibly unnormalized binary distribution, we need to show that the following system of linear equations indeed has a solution $(w^*, b^*)$ for any $p, q > 0$:

$$\begin{cases} w^* \times 1 + b^* = p \\ w^* \times 2 + b^* = q \end{cases} \quad (1)$$

Subtracting the two equations

$$w^* = q - p \quad (2)$$

Substituting $w^*$ in the first equation, and solving for $b^*$

$$b^* = 2p - q \qquad (3)$$

Therefore, $(w^* = q - p, b^* = 2p - q)$ is a solution of the system of linear equations indicated in (1). In fact, $(w^* = q - p, b^* = 2p - q)$ is the unique solution, as the determinant of the coefficient matrix $\begin{bmatrix} 1 & 1 \\ 2 & 1 \end{bmatrix}$ is nonzero. This completes the first part of the proof.

In the case of having an asymmetric sigmoid activation function $y = 1/(1 + e^{-(wx+b)})$, we have

$$\begin{cases} 1/(1 + e^{-(w^* \times 1 + b^*)}) = p \\ 1/(1 + e^{-(w^* \times 2 + b^*)}) = q \end{cases} \qquad (4)$$

which simplifies to

$$\begin{cases} e^{-(w^* \times 1 + b^*)} = \dfrac{1}{p} - 1 \\ e^{-(w^* \times 2 + b^*)} = \dfrac{1}{q} - 1 \end{cases} \qquad (5)$$

and subsequently to

$$\begin{cases} w^* \times 1 + b^* = -\ln\left(\dfrac{1}{p} - 1\right) \\ w^* \times 2 + b^* = -\ln\left(\dfrac{1}{q} - 1\right) \end{cases} \qquad (6)$$

Following the same logic carried out in the case of (1), $(w^* = \ln\left(\frac{1}{p} - 1\right) - \ln\left(\frac{1}{q} - 1\right), b^* = \ln\left(\frac{1}{q} - 1\right) - 2\ln\left(\frac{1}{p} - 1\right))$, which simplifies to $(w^* = \ln\left(\frac{q(1-p)}{p(1-q)}\right), \ b^* = \ln\left(\frac{(1-q)p^2}{q(1-p)^2}\right))$, is the unique solution of the system of linear equations indicated in (4). This completes the whole proof.

**Example:**

**Distribution:** $(1, p; 2, q)$ with p=.8 and q=.2

x=1 (input) → y=.8 (output)

x=2 (input) → y=.2 (output)

neural network output $y = 1/(1 + e^{-(wx+b)})$

$$w^* = \ln\left(\frac{q(1-p)}{p(1-q)}\right) = -2.7726$$

$$b^* = \ln\left(\frac{(1-q)p^2}{q(1-p)^2}\right) = 4.1589$$

**Verification:**

$y = 1/\left(1 + e^{-(w^* \times 1 + b^*)}\right) = .8$

$y = 1/\left(1 + e^{-(w^* \times 2 + b^*)}\right) = .2$

## References

Ackley, D., Hinton, G., & Sejnowski, T. (1985). A learning algorithm for boltzmann machines. *Cognitive Science*, *9*(1), 147–169.

Agrawal, S., & Goyal, N. (2012). Analysis of thompson sampling for the multi-armed bandit problem. *Journal of Machine Learning Research*, *23*, 1–26.

Althaus, N., Gliozzi, V., Mayor, J., & Plunkett, K. (2020). Infant categorization as a dynamic process linked to memory: Infant categorisation linked to memory. *Royal Society Open Science*, *7*(10).

Alvarez, G. (2011). Representing multiple objects as an ensemble enhances visual cognition. *Trends in Cognitive Sciences*, *15*(3), 122–131.

Aslin, R., Saffran, J., & Newport, E. (1998). Computation of conditional probability statistics by 8 month old infants. *Psychological Science*, *9*(4), 321–324.

Baluja, S., & Fahlman, S. (1994). *Reducing network depth in the cascade-correlation learning architecture*. Carnegie Mellon University. Pittsburgh, PA.

Battaglia, P., Hamrick, J., & Tenenbaum, J. (2013). Simulation as an engine of physical scene understanding. *Proceedings of the National Academy of Sciences of the United States of America*, *110*(45), 18327–18332.

Behrend, E., & Bitterman, M. (1961). Probability-Matching in the Fish. *The American Journal of Psychology*, *74*(4), 542–551.

Bhui, R., & Gershman, S. (2018). Decision by sampling implements efficient coding of psychoeconomic functions. *Psychological Review*, *125*(6), 985–1001.

Bonawitz, E., Denison, S., Gopnik, A., & Griffiths, T. (2014). Win-Stay, Lose-Sample: A simple sequential algorithm for approximating Bayesian inference. *Cognitive Psychology*, *74*, 35–65.

Bonawitz, E., Denison, S., Griffiths, T., & Gopnik, A. (2014). Probabilistic models, learning algorithms, and response variability: Sampling in cognitive development. *Trends in Cognitive Sciences*, *18*(10), 497–500.

Cohen, J., McClure, S., & Yu, A. (2007). Should I stay or should I go? How the human brain manages the trade-off between exploitation and exploration. *Philosophical Transactions of the Royal Society B: Biological Sciences*, *362*(1481), 933–942.

Courville, A., Bergstra, J., & Bengio, Y. (2011). A spike and slab Restricted Boltzmann Machine. *Journal of Machine Learning Research*, *15*, 233–241.

Dasgupta, I., Schulz, E., & Gershman, S. (2017). Where do hypotheses come from? *Cognitive Psychology*, *96*, 1–25.

Denison, S., Bonawitz, E., Gopnik, A., & Griffiths, T. (2010). Preschoolers sample from probability distributions. In R. Camtrabone & S. Ohlsson (Eds.), *Proceedings of the 32nd annual conference of the Cognitive Science Society* (pp. 2272–2277). Austin, TX: Cognitive

Science Society.

Denison, S., Reed, C., & Xu, F. (2013). The emergence of probabilistic reasoning in very young infants: evidence from 4.5- and 6-month-olds. *Developmental Psychology*, *49*(2), 243–249.

Denison, S., & Xu, F. (2010). Twelve- to 14-month-old infants can predict single-event probability with large set sizes. *Developmental Science*, *13*(5), 798–803.

Denison, S., & Xu, F. (2014). The origins of probabilistic inference in human infants. *Cognition*, *130*(3), 335–347.

Eckert, J., Call, J., Hermes, J., Herrmann, E., & Rakoczy, H. (2018). Intuitive statistical inferences in chimpanzees and humans follow Weber's law. *Cognition*, *180*, 99–107.

Elman, J. (2005). Connectionist models of cognitive development: Where next? *Trends in Cognitive Sciences*, *9*(3 SPEC. ISS.), 111–117.

Elman, J., Bates, E., Johnson, M., Karmiloff-Smith, A., Parisi, D., & Plunkett, K. (1996). *Rethinking innateness : a connectionist perspective on development*. Cambridge, Mass.: MIT Press.

Félix, M. A., & Barkoulas, M. (2015). Pervasive robustness in biological systems. *Nature Reviews Genetics*, *16*(8), 483–496.

Gerken, L., Balcomb, F., & Minton, J. (2011). Infants avoid "labouring in vain" by attending more to learnable than unlearnable linguistic patterns. *Developmental Science*, *14*(5), 972–979.

Gershman, S. (2018). Deconstructing the human algorithms for exploration. *Cognition*, *173*(August 2017), 34–42.

Gershman, S., Horvitz, E., & Tenenbaum, J. (2015). Computational rationality: A converging paradigm for intelligence in brains, minds, and machines. *Science*, *349*(6245), 273–278.

Greggers, U., & Menzel, R. (1993). Memory dynamics and foraging strategies of honeybees. *Behavioral Ecology and Sociobiology*, *32*(1), 17–29.

Griffiths, T., Vul, E., & Sanborn, A. (2012). Bridging levels of analysis for probabilistic models of cognition. *Current Directions in Psychological Science*, *21*(4), 263–268.

Hastings, W. (1970). Monte carlo sampling methods using Markov chains and their applications. *Biometrika*, *57*(1), 97–109.

Hinton, G., Osindero, S., & Teh, Y. (2006). A fast learning algorithm for deep belief nets. *Neural Computation*, *18*(7), 1527–1554.

Hinton, G., & Salakhutdinov, R. (2006). Reducing the Dimensionality of Data with Neural Networks. *Science*, *313*(July), 504–507.

Kharratzadeh, M., & Shultz, T. (2016). Neural implementation of probabilistic models of cognition. *Cognitive Systems Research*, *40*, 99–113.

Kirk, K., & Bitterman, M. (1965). Probability-learning by the turtle. *Science*, *148*(3676), 1484–1485.

Kitano, H. (2004). Biological robustness. *Nature Reviews Genetics*, *5*(11), 826–837.

Koller, D., & Friedman, N. (2009). *Probabilistic graphical models: Principles and techniques*. Cambridge, MA: MIT press.

Lany, J., & Gómez, R. L. (2008). Twelve-month-old infants benefit from prior experience in statistical learning. *Psychological Science*, *19*(12), 1247–1252.

Lecun, Y., Bengio, Y., & Hinton, G. (2015). Deep learning. *Nature*, *521*(7553), 436–444.

Lieder, F., Griffiths, T., Huys, Q., & Goodman, N. (2018). The anchoring bias reflects rational use of cognitive resources. *Psychonomic Bulletin and Review*, *25*(1), 322–349.

Ma, W., Beck, J., Latham, P., & Pouget, A. (2006). Bayesian inference with probabilistic

population codes. *Nature Neuroscience*, *9*(11), 1432–1438.

Mareschal, D. (2010). Computational perspectives on cognitive development. *Wiley Interdisciplinary Reviews: Cognitive Science*, *1*(5), 696–708.

Mareschal, D., & French, R. (2017). Tracx2: A connectionist autoencoder using graded chunks to model infant visual statistical learning. *Philosophical Transactions of the Royal Society B: Biological Sciences*, *372*(1711).

Mareschal, D., Johnson, M., Sirois, S., Spratling, M., Thomas, M., & Westermann, G. (2007). *Neuroconstructivism, Vol. I: How the brain constructs cognition*. Oxford: Oxford University Press.

Marr, D. (2010). *Vision: a computational investigation into the human representation and processing of visual information*. Cambridge, MA: MIT Press.

Metropolis, N., Rosenbluth, A., Rosenbluth, M., Teller, A., & Teller, E. (1953). Equation of state calculations by fast computing machines. *The Journal of Chemical Physics*, *21*(6), 1087–1092.

Moreno-Bote, R., Knill, D., & Pouget, A. (2011). Bayesian sampling in visual perception. *Proceedings of the National Academy of Sciences of the United States of America*, *108*(30), 12491–12496.

Munakata, Y., & McClelland, J. (2003). Connectionist models of development. *Developmental Science*, *6*, 413–429.

Newell, A. (1994). *Unified theories of cognition*. Cambridge, MA: Harvard University Press.

Nobandegani, A., da Silva Castanheira, K., O'Donnell, T., & Shultz, T. (2019). On Robustness : An Undervalued Dimension of Human Rationality. *Proceedings of the 17th International Conference on Cognitive Modeling*, 1–6.

Nobandegani, A., & Shultz, T. (2017). Converting cascade-correlation neural nets into probabilistic generative models. In G. Gunzelmann, A. Howes, T. Tenbrink, & E. J. Davelaar (Eds.), *Proceedings of the 39th Annual Conference of the Cognitive Science Society* (pp. 1029–1034). Austin, TX: Cognitive Science Society.

Nobandegani, A., & Shultz, T. (2018). Example generation under constraints using cascade correlation neural nets. In T. Rogers, M. Rau, X. Zhu, & C. Kalish (Eds.), *Proceedings of the 40th Annual Meeting of the Cognitive Science Society* (pp. 2385–2390).

Nobandegani, A., & Shultz, T. (2020). A Resource-Rational, Process-Level Account of the St. Petersburg Paradox. *Topics in Cognitive Science*, *12*, 417–432.

Oakes, L., Madole, K., & Cohen, L. (1991). Infants' object examining: Habituation and categorization. *Cognitive Development*, *6*(4), 377–392.

Piaget, J., & Inhelder, B. (1975). *The origin of the idea of chance in children*. New York: Norton.

Poor, H. (1994). *An Introduction to Signal Detection and Estimation* (2nd ed.). New York: Springer-Verlag.

Saffran, J., Aslin, R., Johnson, E., & Newport, E. (1999). Statistical learning of tone sequences by human infants and adults. *Cognition*, *70*(1), 27–52.

Sahani, M., & Dayan, P. (2003). Doubly Distributional Population Codes : *Neural Computation*, *15*, 2255–2279.

Salakhutdinov, R. (2008). *Learning and evaluating Boltzmann machines*. *University of Toronto Machine Learning Technical Report (UTML TR 2008–002)*.

Savin, C., & Denève, S. (2014). Spatio-temporal representations of uncertainty in spiking neural networks. In Z. Ghahramani, M. Welling, C. Cortes, N. D. Lawrence, & K. Q. Weinberger (Eds.), *Advances in Neural Information Processing Systems 27* (pp. 2024–2032). Curran Associates, Inc.

Shultz, T. (2003). *Computational developmental psychology.* Cambridge, MA: MIT Press.

Shultz, T. (2010). Computational Modeling of Infant Concept Learning: The Developmental Shift from Features to Correlations. In L. Oakes, C. Cashon, M. Casasola, & D. Rakison (Eds.), *Infant Perception and Cognition: Recent Advances, Emerging Theories, and Future Directions* (pp. 125–152). New York: Oxford University Press.

Shultz, T. (2012). A constructive neural-network approach to modeling psychological development. *Cognitive Development*, *27*(4), 383–400.

Shultz, T. (2013). Computational models in developmental psychology. In P. D. Zelazo (Ed.), *Oxford handbook of developmental psychology, Vol. 1: Body and mind* (pp. 477–504). New York: Oxford University Press.

Shultz, T. (2017). Constructive artificial neural-network models for cognitive development. In N. Budwig, E. Turiel, & P. D. Zelazo (Eds.), *New Perspectives on Human Development* (pp. 13–26). Cambridge: Cambridge University Press.

Shultz, T., & Bale, A. (2001). Neural network simulation of Infant familiarization to artificial sentences: Rule-like behavior without explicit rules and variables. *Infancy*, *2*(4), 501–536.

Shultz, T., & Bale, A. (2006). Neural networks discover a near-identity relation to distinguish simple syntactic forms. *Minds and Machines*, *16*(2), 107–139.

Shultz, T., & Cohen, L. (2004). Modeling age differences in infant category learning. *Infancy*, *5*(2), 153–171.

Shultz, T., & Doty, E. (2014). Knowing when to quit on unlearnable problems: another step towards autonomous learning. In J. Mayor & P. Gomez (Ed.), *Computational Models of*

*Cognitive Processes* (pp. 211–221). London: World Scientific.

Shultz, T., & Sirois, S. (2008). Computational models of developmental psychology. In R. Sun (Ed.), *The Cambridge handbook of computational psychology* (pp. 451–476). New York: Cambridge University Press.

Spelke, E. (2000). Core knowledge. *American Psychologist*, *55*(11), 1233–1243.

Summerfield, C., & Tsetsos, K. (2015). Do humans make good decisions? *Trends in Cognitive Sciences*, *19*(1), 27–34.

Teglas, E., Vul, E., Girotto, V., Gonzalez, M., Tenenbaum, J., & Bonatti, L. (2011). Pure reasoning in 12-month-old infants as probabilistic inference. *Science*, *332*(6033), 1054–1059.

Vul, E., Goodman, N., Griffiths, T., & Tenenbaum, J. (2014). One and done? Optimal decisions from very few samples. *Cognitive Science*, *38*(4), 599–637.

Westermann, G., Sirois, S., Shultz, T., & Mareschal, D. (2006). Modeling developmental cognitive neuroscience. *Trends in Cognitive Sciences*, *10*(5), 227–232.

Wilson, R., Geana, A., White, J., Ludvig, E., & Cohen, J. (2014). Humans use directed and random exploration to solve the explore-exploit dilemma. *Journal of Experimental Psychology: General*, *143*(6), 2074–2081.

Xu, F., & Garcia, V. (2008). Intuitive statistics by 8-month-old infants. *Proceedings of the National Academy of Sciences*, *105*(13), 5012–5015.

Yu, L., Nobandegani, A., & Shultz, T. (2019). Neural Network Modeling of Learning to Actively Learn. *Proceedings of the 17th International Conference on Cognitive Modeling*, 1–6.

Zemel, R., Dayan, P., & Pouget, A. (1998). Probabilistic interpretation of population codes.

*Neural Computation, 10(2), 403-430.*, *10*, 403–430.